\documentclass{emulateapj}
\usepackage{graphicx}
\usepackage[space]{grffile}
\usepackage{latexsym}
\usepackage{textcomp}
\usepackage{longtable}
\usepackage{multirow,booktabs}
\usepackage{amsfonts,amsmath,amssymb}
\usepackage{url}
\usepackage{hyperref}
\hypersetup{colorlinks=true,pdfborder={0 0 0}}
\newif\iflatexml\latexmlfalse
\usepackage[utf8]{inputenc}
\usepackage[english]{babel}
\usepackage[usenames, dvipsnames]{color}

\usepackage{natbib}

\shorttitle{Driving Solar Giant Cells through Near-Surface Plumes}
\shortauthors{Nelson et al.}

\begin{document}


\title{Driving Solar Giant Cells through the Self-Organization of Near-Surface Plumes}


\author{Nicholas J.~Nelson\altaffilmark{1}, Nicholas A.~Featherstone\altaffilmark{2}, Mark S.~Miesch\altaffilmark{3}, \& Juri Toomre\altaffilmark{4}}
\altaffiltext{1}{Department of Physics, California State University, Chico}
\altaffiltext{2}{Research Computing, University of Colorado}
\altaffiltext{3}{High Altitude Observatory, National Center for Atmospheric Research}
\altaffiltext{4}{JILA \& Department of Astrophysical and Planetary Sciences, University of Colorado}

\email{Contact: njnelson@csuchico.edu\\}




\begin{abstract}
Global 3D simulations of solar giant-cell convection have provided significant insight into the processes which yield the Sun's observed differential rotation and cyclic dynamo action. However, as we move to higher resolution simulations a variety of codes have encountered what has been termed the convection conundrum. As these simulations increase in resolution and hence the level of turbulence achieved, they tend to produce weak or even anti-solar differential rotation patterns associated with a weak rotational influence (high Rossby number) due to large convective velocities. One potential culprit for this convection conundrum is the upper boundary condition applied in most simulations which is generally impenetrable. Here we present an alternative stochastic plume boundary condition which imposes small-scale convective plumes designed to mimic near-surface convective downflows, thus allowing convection to carry the majority of the outward solar energy flux up to and through our simulated upper boundary. The use of a plume boundary condition leads to significant changes in the convective driving realized in the simulated domain and thus to the convective energy transport, the dominant scale of the convective enthalpy flux, and the relative strength of the strongest downflows, the downflow network, and the convective upflows. These changes are present even far from the upper boundary layer. Additionally, we demonstrate that in spite of significant changes, giant cell morphology in the convective patterns is still achieved with self-organization of the imposed boundary plumes into downflow lanes, cellular patterns, and even rotationally-aligned banana cells in equatorial regions. This plume boundary presents an alternative pathway for 3D global convection simulations where driving is non-local and may provide a new approach towards addressing the convection conundrum. 

\end{abstract}%

\bibliographystyle{apj}

\section{Deep Solar Convection} \label{sec:deep}

Guided by both theoretical and observational arguments, we seek to explore the effect of a stochastic plume boundary condition in global solar convection simulations. Past global convective models have generally used impenetrable upper boundary conditions both for numerical simplicity and because of the impressive triumphs of such models in reproducing solar differential rotation \citep{Brun2002,Miesch2006, Gastine_Yadav_Morin_Reiners_Wicht_2014, Fan_Fang_2014, Karak2015}, exploring the strength, topology, and variability of dynamo action \citep{Browning_Miesch_Brun_Toomre_2006, Brown_Browning_Brun_Miesch_Toomre_2010, Racine_Charbonneau_Ghizaru_Bouchat_Smolarkiewicz_2011, Guerrero_Kapyla_2011, Gastine_Duarte_Wicht_2012, Nelson_Miesch_2014}, and the nature of the deep meridional circulation \citep{Featherstone2015, Passos_Charbonneau_Miesch_2015, hotta15}. These successes have all been achieved for simulations that not only do not treat the near-surface layers and the granular and super-granular scales of convection that reside there, but that also do not consider the effects of near-surface flows on the dynamics of the deeper, global-scale convection other than the transport of the solar luminosity. Here we present an alternative implementation of the impact of near surface convection on giant-cell convection in a 3D global solar simulation.

One of the touchstones of global solar convective simulations has been the reproduction of solar-like differential rotation. Simulations maintain solar-like rotational constraints on their convective flows by balancing their inertial and Coriolis forces. In order to achieve the needed strong rotational constraints, modelers have often found it necessary to artificially enhance the dissipation, decrease the luminosity, or increase the rotation rate of their simulations, and thus maintain the fast-equator, slow-pole pattern of differential rotation observed in the Sun. In addition to the clear problems this presents for models which seek to understand the solar differential rotation, this points to a larger question of why global convection models appear to be incorrectly simulating aspects of giant cell convection such as the amplitudes and scales of giant-cell convection, particularly for the highest levels of resolution current possible \citep{OMara2016}.

In this paper we will explore a new upper boundary condition which may provide a pathway towards highly-turbulent solar-like simulations which incorporate additional effects from near-surface convection and produce significantly different giant cell convective flows. These flows are capable of achieving solar-like differential rotation, even at high resolution, using the solar rotation rate and luminosity. We will show that the use of a stochastic plume boundary condition designed to mimic near-surface convective downflows can substantially alter the resulting giant cell convection by changing the nature of the convective driving, shifting the scales at which the solar enthalpy flux is transported, and altering the relative contribution of upflows and downflows. Further, we will show that this can be done while still producing convective giant cells through plume self-organization and that the resulting flows achieve a higher level of rotational constraint than a comparable simulation with an impenetrable upper boundary.  This bodes well for generating the solar differential rotation, which is thought to arise from the Coriolis-induced Reynolds stress arising from such rotationally-constrained giant cells \citep{Miesch2005}.

\subsection{Observational Constraints on Near-Surface and Deep Convection}

Solar convection has long been divided into several distinct spatial scales, including granulation, supergranulation, and giant cell convection \citep{Nordlund_Stein_Asplund_2009, Rieutord2010, Rincon2016}.  Granular and supergranular convection dominate the near-surface velocity field and have long been observable through direct imaging in the case of granules, and through local helioseismology in the case of supergranules \citep{Gizon_Birch_2005}. Both are also clearly seen in the power spectrum of solar surface velocities \citep{Hathaway2015}. Granulation's characteristic scale of about 1 Mm is the result of rapid radiative cooling at the photosphere. Supergranulation's origin as a preferred scale of convection on the order to 30 Mm is less clear \citep[see][]{Rast2003,Rieutord2010, Featherstone2016b,Rincon2016}. While the both the direction and the existence of a casual connection is debated, supergranules extend to about the same depth as the near-surface shear layer. Together these near-surface convective motions fill roughly the outer 5\% of the solar interior by radius. 

Giant cell convection is here used as the generic term for the flows responsible for transporting angular momentum and energy between the base of the convection zone near $0.72 R_\odot$ and $0.95 R_\odot$. It is likely not a preferred scale in the same sense as either granulation or supergranulation. Giant cell convection may instead peak at a scale which varies continuously with depth. Furthermore giant cell convection likely influences and is influenced by supergranualar flows in the near-surface layers. The giant cell convection is thought to be largely responsible for the global solar dynamo, the generation of sunspots, and the 22-year solar magnetic cycle \citep{Miesch_Toomre_2009, Charbonneau_2010}.

Giant cell convection has proven far more elusive to observations than its near-surface counterparts,  but recent measurements have begun to constrain its scale and amplitude.  The first observational constraint put forward by \citep{Hanasoge_Duvall_DeRosa_2010} used local helioseismic measurements to place an upper limit on the expected large-scale motions of giant cells to amplitudes of about 10 m/s per spherical harmonic mode. Subsequent work tightened these upper limits to about 1 m/s at a depth of about $0.96 R_\odot$ \citep{Hanasoge_Duvall_Sreenivasan_2012}. Meanwhile, \citep{Hathaway_Upton_Colegrove_2013} used supergranulation tracking to detect giant cell flows with rough velocity estimates of about 20 m/s on scales of about 200 Mm or spherical harmonic degree $\ell \approx 20$ though the depth of this measurement was somewhat uncertain. Another detection of giant cells at a scale of about 125 Mm via a derivative of photospheric magnetic field data was reported by \citep{McIntosh_Wang_Leamon_Scherrer_2014}. Most recently \citep{Greer2015} have presented an independent analysis of local helioseismic data claiming a detection of convection flow amplitudes of about 80 m/s at $0.96 R_\odot$ \citep[see also][]{Greer2016, Greer2016a} The disagreement of nearly two orders of magnitude between the various observational inferences is as of yet unresolved. However, once sorted out they should provide valuable observational constraints for modelers of deep solar convection. 
  
\subsection{Advances in Modeling Solar Convection} \label{sec:intro_model}
  
In concert with the advances in observations of giant cell convection, modeling efforts have undergone rapid advances as well. Beginning with the work of \cite{Gilman1983} and \cite{Glatzmaier_1985}, 3D global convection simulations have progressed with roughly the same exponential growth as modern computers. Of special relevance to this work are the studies of convection and differential rotation. \cite{Brun_Toomre_2002} investigated the dependence of simulation parameters such as Prandtl and Reynolds number on the resulting differential rotation produced by simulations with on the order of $2 \times 10^6$ grid points. \cite{Miesch2006} showed that the application of a weak latitudinal entropy gradient on the lower boundary could move the differential rotation from cylindrical to more conical contours, in better agreement with helioseismic measurements \citep{Howe2009}. \cite{Miesch_Brun_DeRosa_Toomre_2008} further explored the nature of giant cell convection as much higher levels of resolution with over $5 \times 10^8$ grid points, however at this resolution a troubling development was noticed. The high resolution, highly turbulent model showed a reduction in the differential rotation contrast in latitude of more than 50\% compared to the much lower resolution version of \cite{Brun_Toomre_2002}. 

This puzzling trend of higher Rayleigh-number simulations producing weaker differential rotation profiles was recently explored by \cite{Gastine_2013} who showed that the transition from strong, solar-like differential rotation through weak differential rotation to anti-solar behavior with rapidly rotating poles and a slowly rotating equator can be addressed simply in terms of the Rossby number, or the ratio of inertial forces to the Coriolis force. \cite{Gastine2013} showed this primarily by modifying the Coriolis forces by changing the bulk rotation rate. \cite{Featherstone2015} produced similar conclusions by modifying the convection and therefore the inertial forces. Such sensitivities have come to be loosely called the `convection conundrum’. There are some indications that magnetic fields may modify the transition point between solar and anti-solar differential rotation \citep[e.g.,][]{Fan_Fang_2014, Mabuchi_Masada_Kageyama_2015, Karak2015, Fan2016}. The need for solar-like simulations to require low Rossby numbers has been expressed in a theoretical argument by \cite{Miesch_Featherstone_Rempel_Trampedach_2012}.

The dependence of differential rotation on the Rossby number given by
\begin{equation}
\mathrm{Ro} = \frac{\mathcal{U}}{2 \Omega \mathcal{L}} \; ,
\end{equation}
where $\mathcal{U}$ and $\mathcal{L}$ are the characteristic velocity and length scales, respectively, and $\Omega$ is the bulk angular velocity of the star. Thus if $\Omega$ is set to the solar value and a simulation does not achieve a solar-like low Rossby number, this can only indicate that the simulated giant cell convection is too small, too fast, or both. This understanding of the essential role of rotational constraint allows the use of differential rotation as a diagnostic of the convective dynamics at play. The simulation with weak differential rotation of  \cite{Miesch_Brun_DeRosa_Toomre_2008} used the same bulk rotation rate, and hence we might expect the Corilois force, denominator in the Rossby number, to be roughly solar-like. The departure from solar behavior must then be the result of a change in the inertial forces, particularly those at the large spatial scales responsible for the transport of angular momentum.

Global-scale convection simulations with solar-like stratification are limited by computational resources to regions well-below the solar photosphere, as the near-surface layers require very high spatial and temporal resolution in addition to additional physics such as full compressibility, a detailed equation of state, and full radiative transfer. Thus global convection models generally are restricted to depths below $0.98 R_\odot$ and use an impenetrable condition on their upper and lower boundaries in which $u_r = 0$. One alternative to this formulation is the use of a modified stratification where an extended layer of convectively stable stratification is added to the top of a global model. \cite{Warnecke2016} have founnd that this formulation results in a further weakening of differential rotation, thus exacerbating the `convection conundrum' further. 

In simulations of giant-cell solar convection, the solar luminosity is introduced into the domain through a physically-accurate radiative flux at the base of the simulation. This flux is taken from solar structure models and is strongly supported by global helioseismic measurements \citep{Christensen-Dalsgaard1996}. In simulations that follow the solar stratification energy must leave the top of the domain via thermal diffusion. In this manner our hydrodynamic simulations resemble a spherical variant of the classic Rayleigh-Benard convection problem which has been studied through theory, simulations, and experiments for close to 90 years \citep{Chandrasekhar_1961,Getling_1998}. The major difference is that in Rayleigh-Benard convection the overall heat flux through the horizontal layer is determined by how the convective flows respond to the boundary layers formed at the upper and lower walls, which are typically maintained at fixed temperatures. Thus in Rayleigh-Bernard convection the convection itself determines the heat flux. In global convective simulations of sun-like stars we demand that a fixed luminosity enter the system at the lower boundary and that the same fixed luminosity exit the system through the upper boundary.

The primary dimensionless parameter that measures convective driving is the Rayleigh number. The Rayleigh number can be defined either as an input to the simulation \citep[e.g.,][]{Featherstone2016b} or as an output of the simulation \citep[e.g,][]{Featherstone2015, Warnecke2016, Yadav2016}. Here let us consider a Rayleigh number based on the output of our simulations, given by
\begin{equation}
\mathrm{Ra} = - \left( \frac{ d \rho }{ d S } \right)_P \frac{ \Delta S g L^3 }{ \rho \nu \kappa } .
\end{equation} 
In this formulation Ra is proportional to the entropy difference  $-\Delta S$ and inversely proportional to each of the diffusion coefficients $\nu$ and $\kappa$. If we want to drive a more turbulent system by lowering $\nu$ and $\kappa$, we should get a correspondingly higher value of the Rayleigh number. We demand that thermal diffusion carry a solar luminosity out of the top of our domain. This provides an additional constraint on $\partial S / \partial r$ at the upper boundary such that
\begin{equation}
 \frac{ \partial S }{ \partial r }  = - \frac{ L_\odot }{ 4 \pi R_o^2 \kappa \bar{\rho} \bar{T} } ,
\end{equation} 
and thus $ \partial S / \partial r \propto \kappa^{-1}$. Further $\Delta S \propto w_b (\partial S / \partial r)$ where $w_b$ is the width of the boundary layer. By lowering $\kappa$ \cite{Featherstone2016} also find there is a nonlinear feedback that yields a narrower upper boundary layer with $w_b \propto \kappa^{1/2}$. Combining all these effects yields a Rayliegh number that is proportional to $\kappa^{-3/2} \nu^{-1}$. Using an alternate, input-based definition of the Rayleigh number \cite{Featherstone2016} and \cite{OMara2016} found that what they term a flux Rayleigh number for these systems is proportional to $\kappa^{-2} \nu^{-1}$. A real physical system such as the solar convection zone would not experience this additional increase in Rayleigh number as it does not rely on a diffusive boundary layer to transport energy. In practice this effect was not significant for the moderate levels of turbulence achieved in previous simulations. However, as resolution has increased the narrowing of the diffusive upper boundary layer has led to a different scaling of the Rayleigh number. These over-driven flows do not feel the proper level of rotational influence and therefore fail to reproduce solar-like differential rotation. 

Given such theoretical challenges, we seek here to explore the effects of a change in the treatment of our upper boundary condition. Specifically, we impose small, short-lived downflow plumes designed to mimic the strongest downflows from supergranular flows which permit our simulation to carry the solar luminosity without forcing a strong diffusive boundary layer at the top of our domain.

\section{Computational Methods}

We present a series of numerical simulations using the Anelastic Spherical Harmonic (ASH) code \citep{Clune_Elliott_Miesch_Toomre_Glatzmaier_1999, Featherstone2015}. ASH solves the compressible equations of hydrodynamics under the anelastic approximation which removes acoustic waves and assumes small perturbations about a background state derived from a stellar structure model \citep{Gough_1969}. Here we will restrict ourselves to models spanning the solar convection zone where the stratification is nearly adiabatic. ASH decomposes the thermodynamic variables into radially varying (spherically symmetric) reference quantities which describe the solar stratification, denoted by overbars, and perturbations around those reference quantities which are functions of all three directions and time. In this work we use a reference state from a 1D stellar structure model designed to match helioseismic measurements \citep[see][]{Howe2009}.  While the reference state is purely adiabatic and constant in time, the perturbations develop a mean entropy gradient which is allowed to evolve.

The anelastic equations used in ASH in this work are described by \cite{Featherstone_Miesch_2015}. Of special note are the standard close-boundary simulations use the impenetrable, stress free boundary conditions given by 
\begin{equation}
u_r = \frac{ \partial }{ \partial r} \left( \frac{u_\theta}{r} \right) = \frac{ \partial }{ \partial r} \left( \frac{u_\phi }{r} \right) = 0 .
\end{equation} 
at both the top and bottom of the domain, along with a constant radial entropy gradient $\partial S / \partial r = 0$ at the bottom and a constant entropy $ S = 0$ at the top. These boundary conditions provide a well-studied set of equations with favorable properties such as conservation of both linear and angular momentum both globally and locally on the boundaries.

\subsection{The Stochastic Plume Boundary Condition}

Conceptually, open boundaries present a number of challenges. On a very basic level an open boundary is an attempt to admit dynamics into a simulation which are not being explicitly treated. Here we make a distinction between truly open, semi-open, and permeable boundary conditions. Truly open boundary conditions allow the resolved interior dynamics of the simulation to control the behavior of the boundaries. They are particularly challenging because they can permit net fluxes of quantities like mass or momentum, removing the global conservation properties of a simulation.

Semi-open boundary conditions are those which are designed to be as open as possible while using minimally invasive techniques to preserve global conservation properties. This may be accomplished using an open boundary condition on the velocity fields while imposing pressure gradients at the boundary to regulate the balance between upflows and downflows, or by applying a volumetric forcing near the boundary to counteract the net fluxes through the boundary. Semi-open boundaries have been used in several codes designed for near-surface solar convection, including MuRAM \citep{Rempel_Schussler_Knolker_2009,Cheung_Rempel_Title_Schussler_2010}, Stagger \citep{Trampedach_Stein_2011}, and CSS \citep{Augustson_Rast_Trampedach_Toomre_2011}. In these codes the bottom boundary condition is open rather than the upper boundary as we are considering for ASH. As the strongest driving in solar and stellar convection is generally believed to occur at the photosphere, opening the lower boundary in a near-surface model may be less problematic than opening the upper boundary in a global convective model. Additionally these are all finite difference codes. Stagger, for example, applies open boundary conditions on outflows but imposes flows and thermodynamic fields on inflows in order to preserve the global conservation of momentum and mass, as well as the solar energy flux \citep{Trampedach_Stein_2011}.

The final category of open boundary conditions is what we term permeable boundary conditions. These models impose flows through their boundaries in a specified way. They may be constant in time or they may vary. A permeable boundary permits flows to enter or exit the domain, but only in a specified manner. Permeable boundaries can easily control the fluxes of conserved quantities since the fluxes are specified with the boundary condition. 

Our plume boundary condition in ASH is composed of three permeable boundary conditions and one semi-open condition. In ASH we impose small-scale plumes of radial velocity and entropy as a time-dependent boundary condition. We also impose a condition on the opening angle of the plumes in order to mitigate pressure perturbations. Finally we impose a semi-open boundary condition designed to permit the internal dynamics of the simulation to set the differential rotation profile on the boundary and use a volumetric torque on the near-boundary layers to enforce conservation of angular momentum. The details of our plume boundary condition are described in Appendix A.

\begin{deluxetable}{lccc}[t]
  \tabletypesize{\footnotesize}
    \tablecolumns{3}
    \tablewidth{0pt}  
    \tablecaption{Values for Plume Parameters
    \label{table:plume_params}}
    \tablehead{  &  
      \colhead{Value Range} &
      \colhead{Units} &
      \colhead{Correlation}
   }
   \startdata
$\mathcal{V}$ & $\left[240, 360 \right] $ & m s$^{-1}$ & -- \\
$\mathcal{E}$ & $\left[ 1.28, 1.56 \right]$ & $10^5$ erg K$^{-1}$ g$^{-1}$ & 0.5 \\
$\delta$ & $\left[ 0.08, 0.12 \right]$ & rad. & 0.5 \\
$\delta R_\mathrm{top}$ & $\left[54.6, 82.0 \right]$ & Mm & 0.5 \\
$\tau$ & $\left[ 12.0, 18.0 \right]$ & days & 0.5 \\
$\theta_p$ & $\left[ 0, \pi \right]$ & rad. & 0 \\
$\phi_p$ & $\left[ 0, \pi / 2 \right]$ & rad. & 0 \\
\enddata
\tablecomments{Plume boundary parameters used where $\mathcal{V}$ is the peak downflow velocity, $\mathcal{E}$ is the peak entropy perturbation, $\delta$ is the plume's angular radius, $\tau$ is the plume lifetime, and $\theta_p$ and $\phi_p$ give the coordinate location of the center of the plume on the outer boundary. Also given is the plume width given by $\delta R_\mathrm{top}$ in Mm for ease of comparison. Correlations are expressed with respect to $\mathcal{V}$.}
\end{deluxetable} 

\begin{figure*}[t]
\begin{center}
\includegraphics[width=0.8\textwidth]{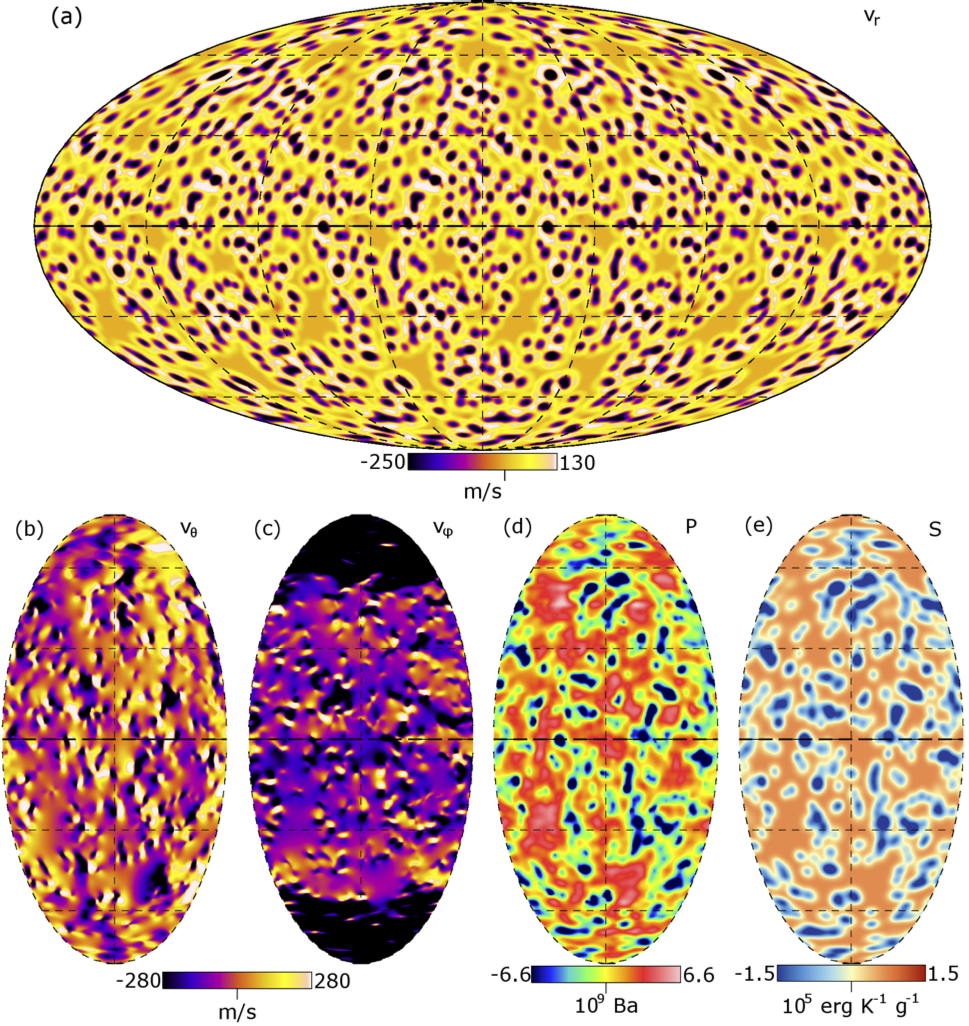}
\caption{Snapshot of the boundary condition applied to case P. (a)~Imposed radial velocity boundary condition with 400 small-scale plumes applied over $90^\circ$ in longitude, shown here repeated four times due to the four-fold periodicity of this simulation. (b)~Latitudinal velocity field at the same instant, shown without the four-fold repetition. (c)~Longitudinal velocity field at the same instant. Polar vortices are visible at high latitude due a combination of the imposed neagtive torque removing angular momentum at high latitudes and as a consequence of the imposed periodicity. (d)~Pressure field at the same instant implicitly set by the convergence of the plumes. (e)~Entropy at the same instant. Plumes are dominated by cold downflows. \label{fig:snapshotBC}%
}
\end{center}
\end{figure*}

Our plume boundary conditions can in principle be used to impose any number of plumes with any range of sizes, positions, and intensities in their downward (or upward) velocities and entropy deficits (or excesses). In practice our choices are limited by (a) the resolution of our simulation, (b) the desire to transport a specified luminosity through the boundary, and (c) our limited knowledge of plume dynamics at $0.98 r_\odot$.  For case P we choose to use 400 plumes on the $90^\circ$ wedge, which is equivalent to 1600 plumes over the full spherical surface. The number of plumes is fixed in time such that when a plume expires it is immediately replaced with a new randomly generated plume at a new location with new properties. Each of our 400 plumes require six parameters, namely the peak downflow velocity $\mathcal{V}$, the peak entropy perturbation $\mathcal{E}$, the angular radius $\delta$, the plume lifetime $\tau$, and the plume center $\left[ \theta_p, \phi_p \right]$. This gives a a total of 2400 parameters to choose. The choice of $\mathcal{V}$ and $\mathcal{E}$ are globally constrained such that we transport the one solar luminosity per spherical shell in enthalpy flux at the outer boundary. Correlations between some of these parameters for a given plume could be expected. For example, one might expect faster plumes to be larger in size, or smaller plumes to have shorter lifetimes. We chose to correlate the velocity amplitude, entropy amplitude, plume width, and plume lifetime. Thus each plume is randomly assigned each of the six parameters from a specified range with the given correlations. In addition, we choose to advect longitudinal position of the plume centers $\phi_p$ with the axisymmetric component of the longitudinal velocity. 

Table~\ref{table:plume_params} lists the range of values, units, and correlation with respect to the velocity amplitude used in case P. For each new plume, whether at the start of the simulation or when a plume is reinitialized after exceeding its assign lifetime, all six parameters are randomly reassigned. The location of the new plume is chosen such that the plume has a uniform probability of appearing at any location on the outer boundary. The amplitude of the velocity perturbation $\mathcal{V}$ is chosen from a uniform random distribution in the range specified. All other variables are chosen randomly from a distribution that is, on average, correlated at the specified level with $\mathcal{V}$. Here we have chosen an arbitrary correlation coefficient of 0.5 between a plume's velocity amplitude and its entropy amplitude, lifetime, and size, however we anticipate near-surface models may be able to provide more realistic choices in the future. Note that this correlation is not a spatial correlation, but rather a correlation over the parameters of the 400 imposed plumes. The value ranges for velocity and entropy perturbation are essentially unconstrained by the numerical properties of our simulations, however the value ranges for plume size and lifetime are here chosen to be as small as permitted given our simulation resolution and time step, as the plumes must be well-resolved both spatially and temporally for numerical stability. In addition, we are constrained by the potential accumulation of pressure perturbations excited by discontinuities in the pressure field due to our plumes. In our formulation, the pressure field is implicitly specified by the radial velocity and its first, second, and third derivatives. Instead of over-specifying our system with additional boundary conditions, we choose instead to allow our plumes to enter the domain out of pressure equilibrium with their surroundings. Within our anelastic framework that does not permit acoustic waves, the resulting pressure perturbations can only be dissipated by coupling to thermal and viscous diffusion or driving bulk flows. 

It is important to note that the plumes used in the simulations reported here are as small and short-lived as computationally feasible, but they are still considerably longer-lived and larger in scale than suggested by near-surface models and helioseismic data. Our plume structures are designed to mimic strong downflows at the interstices of supergranular downflow lanes. They have a mean physical radius of over 68 Mm and a mean lifetime of 15 days. Observational data suggests that supergranules have diameters and lifetimes on the order of 30 Mm and 1.5 days, respectively \citep{Rieutord2010}. While our plumes are likely still too large and long-lived, there is a clear trend towards longer lifetimes and increased spatial scales at greater depth \citep{Spruit1990}.  We anticipate further study at even higher resolution in order to explore the effects of even smaller and shorter-lived plumes than those considered in this work.

Figure~\ref{fig:snapshotBC} shows snapshots of the velocity, pressure, and entropy fields applied at the same instant for case P. While all plumes have the same shape, plume locations are uniformly random on the spherical surface and any overlap between plumes is additive, leading to the surface shown in radial velocity and entropy. The pressure field is implicitly influenced by our choice of the plume convergence parameter, however generally cool downflows are seen to correlate with low pressure regions. The horizontal components of velocity show strong variability, with $u_\phi$ in particular showing strong influence from the interior of the simulation domain. The longitudinal velocity in particular develops strong polar vortices, to which we will return in our discussion of angular momentum conservation.

\subsection{Angular Momentum and Differential Rotation}

\begin{deluxetable*}{lccccccccccc}[t]
  \tabletypesize{\footnotesize}
    \tablecolumns{8}
    \tablewidth{0pt}  
    \tablecaption{Overview of Simulations
    \label{table:overview}}
    \tablehead{\colhead{Case}  &  
      \colhead{$N_r,N_\theta,N_\phi$} &
      \colhead{$\ell_\mathrm{max}$} &
      \colhead{$t_\mathrm{tot}$} &
      \colhead{$t_\mathrm{conv}$} &
      \colhead{Ra} &
      \colhead{Re} &
      \colhead{Pe} &
      \colhead{Ro} &
      \colhead{Roc} &
      \colhead{$\nu$}
   }
   \startdata
 C1 & 200, 256, 128 & 170 & 3840 & 52.0 & $5.33 \times 10^5$  & 93.9 & 23.4 & 0.620 & 1.06 & 24.1 \\
 C2 & 300, 512, 256 & 341 & 6790 & 27.2 & $1.02 \times 10^7$  & 215 & 53.8 & 1.01 & 1.87 & 12.0 \\
 C3 & 500, 1024, 512 & 682 & 10260 & 22.3 & $ 1.54 \times 10^8$ & 819 & 205 & 2.86 & 3.63 & 6.02 \\
 P & 500, 1024, 512 & 682 & 3940 & 25.8 & $ 3.90 \times 10^5$ & 706 & 176 & 2.76 & 0.221 & 6.02
 \enddata
\tablecomments{Computational and fluid parameters for all simulations. 
    The computational resolution is given by the number of radial, latitudinal, and longitudinal grid points ($N_r$, $N_\theta$, and $N_\phi$, respectively), and the maximum spherical harmonic degree $\ell_\mathrm{max}$.
    All simulations rotate at the bulk solar rotation rate of $2.6 \times 10^{-6}$ rad. s$^{-1}$.
        All simulations have inner radius 
	$r_\mathrm{bot} = 5.00 \times 10^{10}$cm and outer radius of 
        $r_\mathrm{top} = 6.83 \times 10^{10}$cm, with 
	$L = (r_\mathrm{top}-r_\mathrm{bot}) = 1.83 \times 10^{10}$cm
	the thickness of the spherical shell.
	The total simulation time $t_\mathrm{tot}$ and the convective timescale ($t_\mathrm{conv} = L/u_\mathrm{rms}$) are given in days.
	The Rayleigh number $\mathrm{Ra} = (-\partial \rho / \partial S)
	\Delta S \, g L^3/\rho \nu \kappa$ uses the entropy difference between the upper and lower boundaries $\Delta S$ (which is determined by the convection) and the diffusion coefficients $\nu$ and $\kappa$ at mid-convection zone.
	Evaluated at $0.85 R_\odot$ are the
	RMS Reynolds number $\mathrm{Re}  = u_\mathrm{rms} L /\nu$, 
	the Peclet number $\mathrm{Pe} = \mathrm{Re} \mathrm{Pr}$,
	the Rossby number $\mathrm{Ro} = \omega / 2 \Omega_0$ where $\omega$ is the RMS vorticity,
	and the convective Rossby number 
	$\mathrm{Roc} = (\mathrm{Ra}/\mathrm{Ta} \, \mathrm{Pr})^{1/2}$.
	Here $u_\mathrm{rms}$ is based on the fluctuating velocity $u'$, which has the axisymmetric
        component removed.
	For all simulations, the Prandtl number $\mathrm{Pr} = \nu / \kappa$ is 0.25.  
	The viscosity $\nu$ is
	quoted at mid-depth (in units of $10^{11}~\mathrm{cm}^2 \; \mathrm{s}^{-1}$).}
\end{deluxetable*} 
  
The use of impenetrable, stress-free boundary conditions in case C1, C2, and C3 automatically provides for the conservation of angular momentum both locally and globally. The radial flux of angular momentum at the boundary is given by 
\begin{equation}
\mathcal{F}_r = \bar{\rho} r \sin{\theta} \left[ u_r u_\phi + r \sin{\theta} \Omega_0 u_r - \nu r \frac{ \partial }{ \partial r} \left( \frac{ u_\phi }{ r } \right)  \right] .
\end{equation}
Clearly, if $u_r = \partial / \partial r \left( u_\phi / r \right) = 0$ as in cases C1, C2, and C3, the radial angular momentum flux is identically zero at each point on the boundary. If not this flux produces a net torque $\tau_B$ given by
\begin{equation}
\tau_B = \frac{1}{r^2} \frac{ \partial }{ \partial r } \left( r^2 \mathcal{F}_r \right) .
\end{equation}

The use of the plume boundary conditions described above does not assure either local or global conservation of angular momentum, allowing the system to spin up or spin down over time.  In practice, the plume boundary condition can yield net fluxes of up to 5\% of the total angular momentum in the convection zone per year. This radial flux is generally outward, leading to a net loss of angular momentum relative to a non-rotating frame of reference. For simulations which can require several simulated years to equilibrate, this change of the global angular momentum is unacceptable and must therefore be mitigated somehow. We choose to do so using a volumetric torque applied to counteract the torque applied by the flows crossing the upper boundary. We chose this corrective torque $\tau_C$ to have the form
\begin{align}
\tau_C = & \left( C_1 \sin \theta + C_2 \sin^2 \theta \right) \times \nonumber \\
& \left( \sqrt{\frac{2}{\pi}} \frac{ \Delta_\tau }{ r^2 } e^{- \left( r - R_o \right)^2 / \left( 2 \Delta_\tau^2 \right) } \right) ,
\end{align}
where $C_1$ and $C_2$ are coefficients chosen to match the time-averaged latitudinal profile of of the flow, and $\Delta_\tau$ is here chosen to be 18.3 Mm or one-tenth of the radial extent of the domain. Note that the radial portion of $\tau_C$ integrates to one. At each time step we preform a Levenberg-Marquardt least-squares fit of $\langle \tau_B \rangle$ to the function $\mathcal{C}_1 \sin \theta + \mathcal{C}_2 \sin^2 \theta$. The coefficients $C_1$ and $C_2$ are then updated by $C_n (t_i) = C_n (t_{i-1}) + \mathcal{C}_n / t_\tau$, where we have chosen $t_\tau = 90$ days or six times the average plume lifetime. This avoids instability due to sharp variations in $\tau_B$ as plume configurations change. This does not, however, assure instantaneous global conservation of angular momentum. In practice the global angular momentum of the system $\mathcal{L}_\mathrm{Tot}$ can wander away from the initial value. We limit those changes to less than 0.1\% of the total angular momentum of the convection zone in a non-rotating frame. The latitudinal coefficients are ``nudged'' upward or downward by 1\% each timestep when the total deviation in global angular momentum is greater than 0.1\%. This nudging alters the corrective torque, and thus the time rate of change of the global angular momentum. This tends to cause $\mathcal{L}_\mathrm{Tot}$ to vary between $0.999 \mathcal{L}_{CZ}$ and $1.001 \mathcal{L}_{CZ}$ rapidly at the beginning of the simulation and then to see the variation period steadily increase as the simulation finds a steady-state.  

\begin{figure}[t]
\begin{center}
\includegraphics[width=0.95\columnwidth]{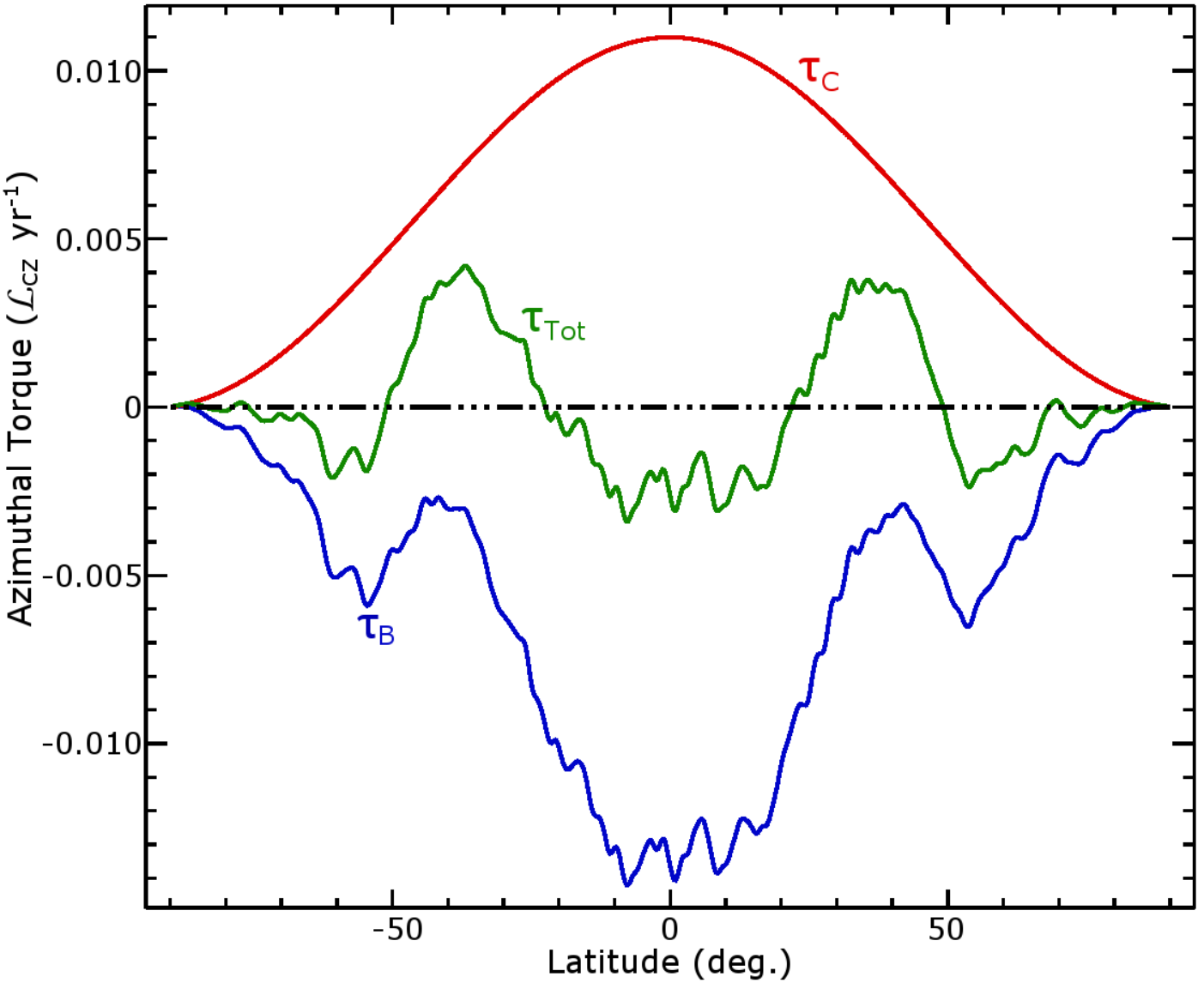}
\caption{Time-averaged torques applied on the domain as a function of latitude by the plume boundary condition $\tau_B$ (blue line), the corrective volumetric torque $\tau_C$ (red line), and the net torque on the system $\tau_\mathrm{Tot}$ (green line) in units of the total angular momentum of the convection zone $\mathcal{L}_{CZ}$ per year. While the total angular momentum of the simulation is constant in time to less than 0.1\%, significant latitudinal redistribution of angular momentum results, yielding deceleration below $20^\circ$, acceleration between about $20^\circ$ and $50^\circ$, and again deceleration at high latitudes. \label{fig:AmomFlux}%
}
\end{center}C
\end{figure}

\begin{figure*}[t]
\begin{center}
\includegraphics[width=0.75\textwidth]{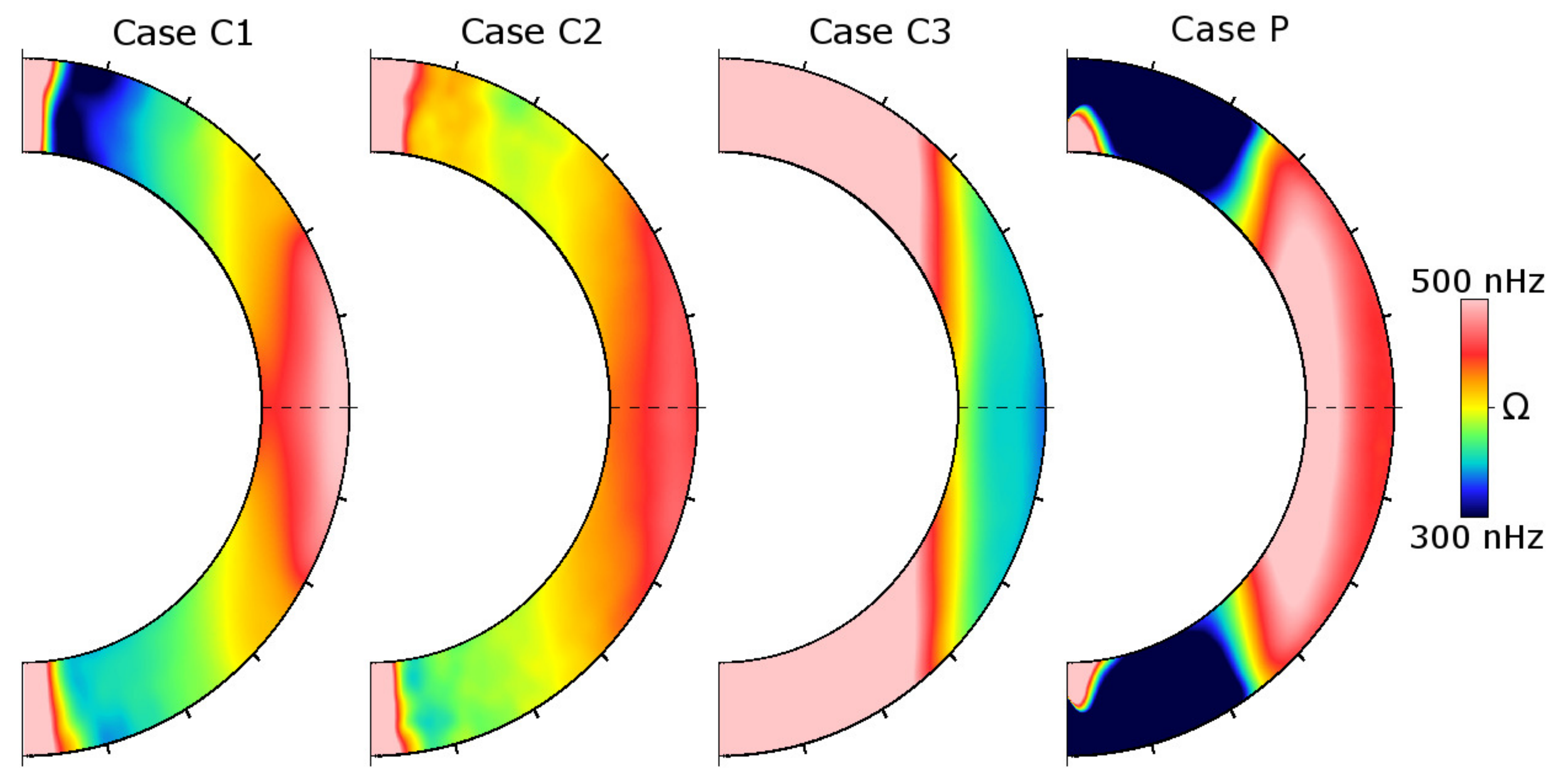}
\caption{Time-averaged differential rotation profiles for cases C1, C2, C3, and P. Cases with closed boundaries move from solar-like differential rotation to anti-solar behavior as the diffusive upper boundary layer becomes progressively thinner. Case P, however, has solar-like behavior at the same resolution as case C3. Despite the overall solar-like behavior, case P shows evidence of the mismatch between the corrective torque and the unconstrained flux of angular momentum through the outer boundary, notably in the polar vortices and the negative radial gradient at the equator. \label{fig:DiffRot}%
}
\end{center}
\end{figure*}

Our choice of function form for $\tau_C$ is motivated by both a theoretical and a practical consideration. It can be reasonably expected that strong downflows from near-surface regions will experience significant deflection by the Coriolis force, leading to a net inward transport of angular momentum at low-latitudes \cite{Miesch2011}. Our plumes are essentially radially directed, so the corrective torque is designed to capture this effect. Additionally we expect the Sun's angular momentum distribution to be in an essentially steady state such that there should be very little net flux through any horizontal surface. This is achieved for ASH simulations with a closed upper boundary condition, reinforcing our expectation. Without a corrective torque, however, we find that our models exhibit a non-trivial net flux of angular momentum which persists over at least hundreds of convective timescales, resulting in an ever-increasing deficit or excess of angular momentum as the simulation evolves. By contrast simulations using the corrective torque approach a statistically stationary state. In case P the net flux of angular momentum time averaged over 100 days was $0.00049 \mathcal{L}_{CZ}$ per year after 2000 days of temporal evolution. All analysis presented in this paper is conducted once this stationary state had been achieved. 

Though the form of the applied torque may capture some realistic aspects of the coupling between the deep convection zone and the surface layers, we acknowledge that it is artificial.  For that reason we focus mainly on the nature of the convective heat transport in these simulations, as opposed to the differential rotation profile that is ultimately achieved.

Figure~\ref{fig:AmomFlux} shows the behavior of $\tau_B$ and $\tau_C$ over a long time-average of 742 days after the differential rotation profile of case P had settled into a steady state. The units plotted are given in the total initial angular momentum of the convection zone $\mathcal{L}_{CZ}$ per year. Even averaging over approximately 50 generations of plumes, $\tau_B$ is still somewhat noisy, however it does show a roughly symmetrical shape about the equator with local minima at about $60^\circ$ and at the equator. This mismatch between the torque due to angular momentum losses and the corrective torque produces significant latitudinal redistribution of angular momentum by speeding up the mid-latitudes while slowing down the polar and equatorial regions.

While our choice of latitudinal profile for $\tau_C$ is at least reasonable under considerations from near-surface models, we cannot rule out other possible torque profiles. The response of our simulation to the corrective torque profile seems to indicate a possible dependence on higher powers of $\sin \theta$. In addition our choice of mass-conserving plumes uniformly distributed over the boundary does not permit a meridional circulation through our outer boundary, which likely plays a significant role in the Sun. Thus this boundary treatment should be regarded as a demonstration of the existence of a potentially useful, albeit non-unique, stochastic boundary condition.

\subsection{Overview of Simulations}

To study the effects of our plume boundary formulation, we present four ASH simulations. Table~\ref{table:overview} presents some relevant computational and nondimensional parameters for these simulations. All four cases extend from $0.72$ to $0.983 R_\odot$ using a realistic solar stratification covering five density scale heights between the upper and lower boundary. To minimize computational expense, all four simulations are conducted in a $90^\circ$ wedge with periodic boundaries in longitude.  Cases C1, C2, and C3 use a standard closed upper boundary condition. Case C1 is chosen to mimic case AB2 of \cite{Miesch2006} but with approximately two additional density scale heights. Case C2 is identical to case C1 but with a reduction in viscosity and thermal diffusivity, and corresponding increase in resolution, by a factor of two. Case C3 is identical to case C2 except for an additional reduction in viscosity and thermal diffusivity by a factor of two, along with a corresponding increase in resolution. Case P is identical to case C3 in every respect expect for the upper boundary condition, which is the plume boundary condition described above. 

Examining the non-dimensional parameters reported in Table~\ref{table:overview} a few trends emerge. First, there is generally a strong increase in the Rayleigh number moving from case C1 to case C3, as would be expected, however case P which has the same diffusion coefficents as case C3 shows a significant decrease in the Rayleigh number due to dramatic decrease in $\Delta S$ in the upper boundary layer and throughout the convection zone. This trend is echoed in the convective Rossby number, however the Rossby number based on the ratio of convective and global vorticities shows equally low levels of rotational constraint for cases C3 and P. The Reynolds number also shows a clear trend with the viscosity even when the plume boundary condition is considered.

\begin{deluxetable}{lcccccc}[t]
  \tabletypesize{\footnotesize}
    \tablecolumns{9}
    \tablewidth{0pt}  
    \tablecaption{Volume-Averaged Energy Densities and Differential Rotation Rates
    \label{table:KEs}}
    \tablehead{\colhead{Case}  &  
      \colhead{TKE} &
      \colhead{DRKE} &
      \colhead{MCKE} &
      \colhead{CKE} &
       \colhead{$\Delta \Omega_{60}$}
   }
   \startdata
 C1 & 36.5 & 25.5 & 0.4 & 10.6 & 131.6  \\
 C2 & 40.3 & 6.3 & 1.0 & 33.0 & 67.2 \\
 C3 & 347.4 & 305.8 & 1.8 & 39.7 & -282.7 \\
 P & 164.9 & 102.8 & 3.1 & 59.0 & 249.6 

\enddata
\tablecomments{For each case TKE gives the total kinetic energy averaged over the simulated volume and in time, DRKE gives the time-averaged kinetic energy in the axisymmetric differential rotation, MCKE gives the time-averaged kinetic energy in the axisymmetric meridional circulation, and CKE gives the remaining kinetic energy in the domain. All kinetic energies are given in units of $10^5$ erg cm$^{-3}$. The time-averaged differential rotation measured at the outer boundary of the simulations between the equator and $\pm 60^\circ$ latitude $\Delta \Omega_{60}$ with negative values indicating anti-solar differential rotation is given in units of nHz. }
\end{deluxetable}
  
Table~\ref{table:KEs} shows the volume-averaged components of the kinetic energy in each of the four simulations. We decompose the kinetic energy of the simulation into three components: differential rotation kinetic energy (DRKE), meridional circulation kinetic energy (MCKE), and convective kinetic energy (CKE), which sum to the total kinetic energy. We define these as
\begin{equation}
\mathrm{TKE} = \frac{1}{V} \int \frac{1}{2} \bar{\rho} \left( u_r^2 + u_\theta^2 + u_\phi^2 \right) \; dV
\end{equation}
\begin{equation}
\mathrm{DRKE} = \frac{1}{V} \int \frac{1}{2} \bar{\rho} \langle u_\phi \rangle^2 \; dV
\end{equation}
\begin{equation}
\mathrm{MCKE} = \frac{1}{V} \int \frac{1}{2} \bar{\rho} \left( \langle u_r \rangle^2 + \langle u_\theta \rangle^2 \right) \; dV
\end{equation}
\begin{equation}
\mathrm{FKE} = \mathrm{TKE} - \mathrm{DRKE} - \mathrm{MCKE}
\end{equation}
where angle brackets denote longitudinal averages.

Figure~\ref{fig:DiffRot} shows the time-averaged differential rotation profiles for all four cases. All four cases here show polar vortices of either extremely rapid or extremely slow rotation. This is largely due to our use of four-fold periodic domains in longitude, which in ASH extend all the way to the pole but do not permit flows to cross the poles. The inability of periodic models to resolve flows over the poles results in anomalous accelerations or decelerations which then cause a roughly $10^\circ$ region ($25^\circ$ in case P) in latitude to effectively decouple from the rest of the domain. Previous work has shown that high-speed vortices can occur in closed boundary simulations using periodic symmetry \cite[see][]{Miesch2006}. The anomalously slow polar vorticies in case P are likely also related to the net negative torque applied by our treatment of angular momentum in the plume boundary model at high latitudes. 

Case C1, like case AB2 of \cite{Miesch2006}, yields strong solar like differential rotation. Case C2 sees a weakening of that profile while case C3 has instead created a strong anti-solar profile. Case P, which is identical to case C3 except for the plume boundary condition, returns to a solar-like differential rotation but with irregularities which can be attributed to the mismatches in the latitudinal profiles of $\tau_B$ and $\tau_C$. Specifically, case P shows very slow poles ($\approx 100$ nHz) and a negative radial gradient of $\Omega$ at the equator due to the net torque on the boundary there being negative, while the mid-latitudes are artificially accelerated. It is possible that these irregularities may be corrected by using a higher-order polynomial in $\sin \theta$ for the latitudinal dependence of $\tau_C$. In spite of the latitudinal redistribution of angular momentum caused by the boundary, it is encouraging that case P builds a strong solar-like differential rotation at high resolution.

We caution that with significant imposed torques from both the plume boundary condition and the corrective volumetric torque, it is difficult to assess the differential rotation profile of this simulation. With this novel plume boundary condition we have achieved solar-like differential rotation with the full solar luminosity at the solar rotation rate at a Reynolds number far beyond those in previous ASH simulations, but our treatment of angular momentum is not self-consistent and includes effectively non-local transport. It is encouraging but far from conclusive that this solar-like differential rotation is maintained in spite of our net torque removing angular momentum at the equator and poles, and adding it at mid-latitudes. In future work we hope to create ASH simulations coupled with near-surface models which can self-consistently treat the flux of angular momentum both into and out of the spherical shell considered here.\\
  
\section{Modification of Convective Energy Transport}

It has long been known that boundary layers play a key role in convective energy transport through the bulk of the domain \citep{Chandrasekhar_1961, Ahlers2009}, however the role of boundary layers and their influence on convective dynamics in rotating, stratified systems is receiving renewed attention \citep{ Featherstone2016, OMara2016, Warnecke2016} Here we compare the convective transport of the solar luminosity in boundary-driven convection with that of plume-driven convection. Clear differences between the closed and plume boundary simulations are seen in the relative balance between convective and diffusive transport, the dominant scales of energy transport, and the roles of the downflow network, strong downward plumes, and upflows. \\

\subsection{Changes in Mean Energy Transport}

Cases C1, C2, and C3 show very different kinetic energies in both their axisymmetric and non-axisymmetric components even though the convective driving at the base of layer is identical in both magnitude and mechanism. Generally, we observe that moving from case C1 to C3 we see gradual reductions in the role of thermal diffusion compensated by a net increase in the total convective transport. We find that enthalpy fluxes increase while kinetic energy fluxes become more negative at a slightly slower rate, yielding an overall increase in convective luminosities. The cause of these changes is clearly the modification of the upper boundary layer.

 \begin{figure}[t]
\begin{center}
\includegraphics[width=0.9\columnwidth]{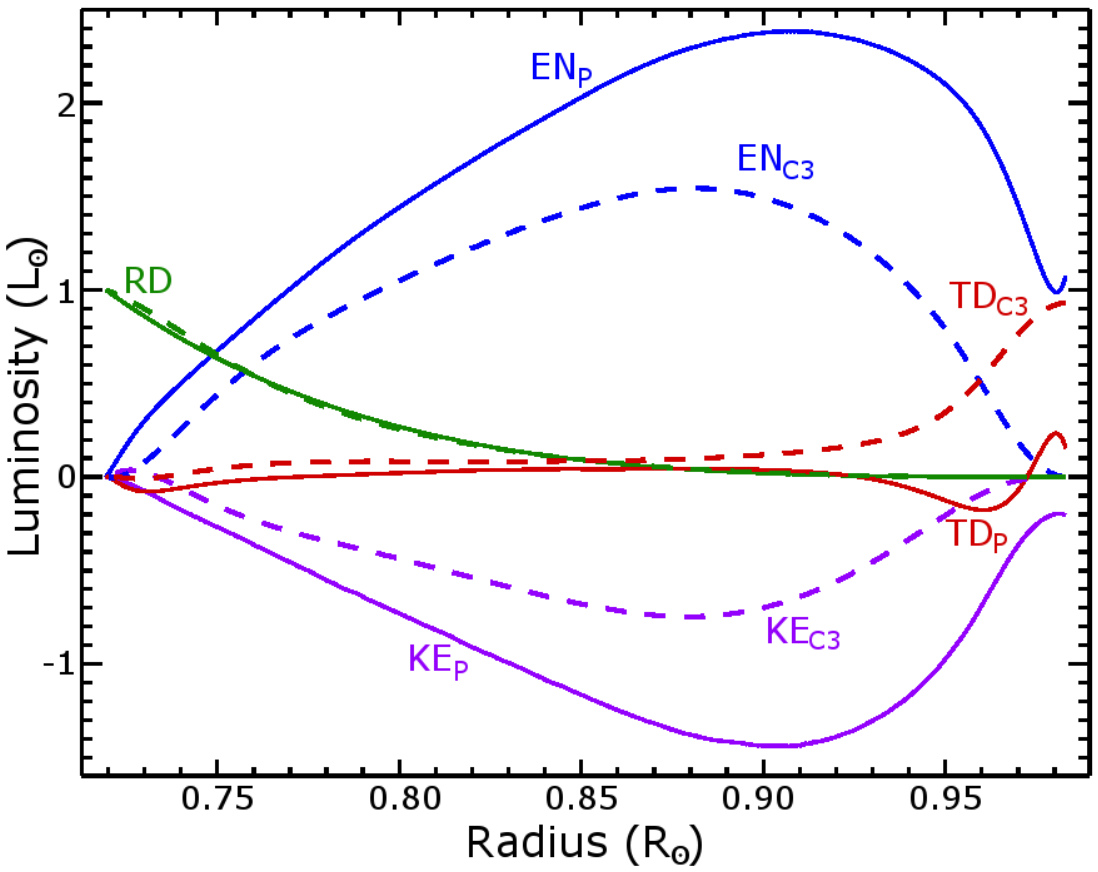}
\caption{Time-averaged luminosity balance for cases C3 (dashed lines) and P (solid lines) taken over 742 days. Contributing to the net energy flux are outward enthalpy luminosity (EN, blue), inward kinetic energy luminosity (KE, purple), radiative luminosity based on solar opacities (RD, green), and resolved thermal diffusion(TD, red). Thermal diffusion is always positive for case C3 indicating that the simulation shows a weak superadiabatic gradient through the bulk of the domain which increases sharply near the upper boundary, while case P shows a slightly negative contribution by thermal diffusion for portions of the domain, indicating a weakly subadiabatic gradient which prevails until about $0.97 R_\odot$. \label{fig:LumBal}%
}
\end{center}
\end{figure}

A dramatic change in the upper boundary condition should be expected to have significant impacts on the bulk convective energy transport, though it is difficult to predict \textit{a priori} what those impacts will be. It may be expected that a plume boundary condition would generally reduce the super-adiabatic gradient at the top of the domain, but the effects are less predictable for the bulk of the domain. By going to a plume boundary condition, we observe significant changes to the net radial transport of energy by convection, the scales which dominate that transport, and the relative roles played by the upward and downward plumes though the bulk of the domain. 

To examine the net radial transport of energy, we decompose the solar energy flux into distinct physical processes which include radiative diffusion $F_\mathrm{RD}$, resolved thermal and viscous diffusion $F_\mathrm{TD}$, enthalpy transport $F_\mathrm{EN}$, and kinetic energy transport $F_\mathrm{KE}$ following \citep{Brun_2004}. In equilibrium, these sum to the solar luminosity when integrated over a spherical surface. They are individually defined as
\begin{equation}
F_\mathrm{RD} = - \kappa_r \bar{\rho} c_P \frac{ d \bar{T} }{ d r }
\end{equation}
\begin{equation}
F_\mathrm{TD} = - \kappa \bar{\rho} \bar{T} \frac{ d \bar{S} }{ d r } - 2 \bar{\rho} \nu \left[ \overline{ \vec{v} \cdot \left( e_{ij} - \frac{1}{3} \left( \nabla \cdot \vec{u} \right) \delta_{ij}\right) } \right]_r 
\end{equation}
\begin{equation}
F_\mathrm{EN} = \bar{\rho} c_P \overline{u_r T} 
\label{eq:enthalpy}
\end{equation}
\begin{equation}
F_\mathrm{KE} = \frac{1}{2} \bar{\rho} \left( \overline{ u_r u^2 } \right)
\end{equation}

Figure~\ref{fig:LumBal} shows the net radial transport of energy due to radiative diffusion, resolved thermal diffusion, kinetic energy transport, and enthalpy transport for cases C3 and P. The quantities plotted are thus the fluxes multiplied by the surface area of each layer. In both cases the energy transported into the bottom of the domain and through its lower third by radiative diffusion is essentially unchanged.  For case C3 only thermal diffusion can transport energy out the top of the domain, while for case P thermal diffusion, enthalpy, and kinetic energy can all play a role. By our choice of plume parameters, the plumes in case P transport $1.05 L_\odot$ of enthalpy flux at the outer boundary and $-0.19 L_\odot$ of  kinetic energy flux. Those values are set by the parameters of the plumes applied on the outer boundary. The flux due to thermal diffusion is then driven to balance the input of a solar luminosity through the lower boundary, here providing $0.15 L_\odot$. Thus over long time averages the the simulation maintains a constant energy flux of one solar luminosity per full spherical shell.

 \begin{figure}
\begin{center}
\includegraphics[width=\columnwidth]{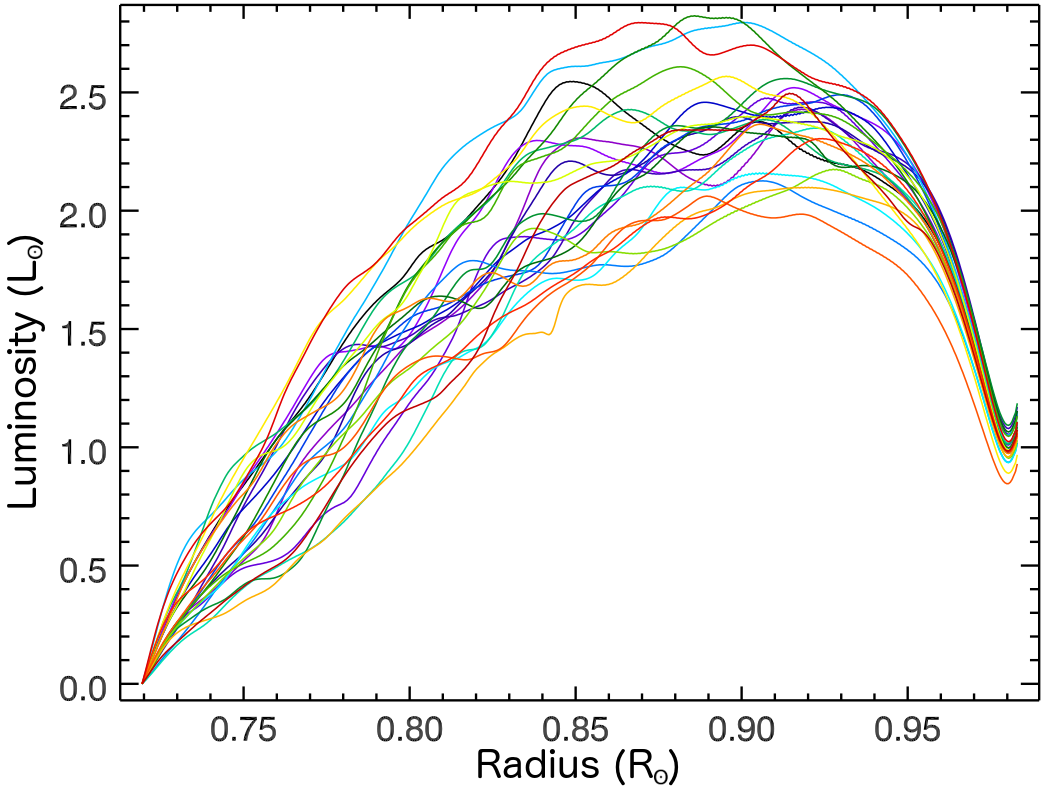}
\caption{Instantaneous snapshots of energy transported by enthalpy $L_\mathrm{EN}$ as a function of radius for case P plotted at 25 times with each snapshot plotted in a different color to highlight the variability in instantaneous energy transport caused by the stochastic plume boundary condition. \label{fig:LumVar}%
}
\end{center}
\end{figure}

Convective energy transport is inherently somewhat variable in time as there is no physical constraint on temporarily heating or cooling a given layer by a convergence or divergence of energy flux. In these models the timescale to adjust the transport by thermal diffusion is very long compared to the timescale for variability in the convective flows, and the diffusive timescale governs the adjustment of the total energy flux leaving the domain in a closed boundary simulation. In a plume boundary simulation, by comparison, considerably more variability is introduced as the plumes and their associated enthalpy and kinetic energy fluxes vary on timescales somewhat shorter than the convective turnover time. Figure~\ref{fig:LumVar} demonstrates this variability in case P by plotting instantaneous snapshots of the radial profile of $L_\mathrm{EN}$ for 25 times. These snapshots sample the interval used to construct the time average seen in Figure~\ref{fig:LumBal}. Short-term variability can yield more than a solar luminosity of variation in enthalpy transport, with variations in kinetic energy flux that are generally anti-correlated but similar in amplitude.

The stochastic nature of the plume boundary condition leads to the internal dynamics of ASH simulations which are inherently more variable in time. This variability is both of greater amplitude and occurs on faster timescales than the variations in similar closed boundary simulations. Much as with the conservation of global angular momentum, a constant net flux of energy through case P is only achieved in a temporally-averaged sense.

 \begin{figure}[t]
\begin{center}
\includegraphics[width=\columnwidth]{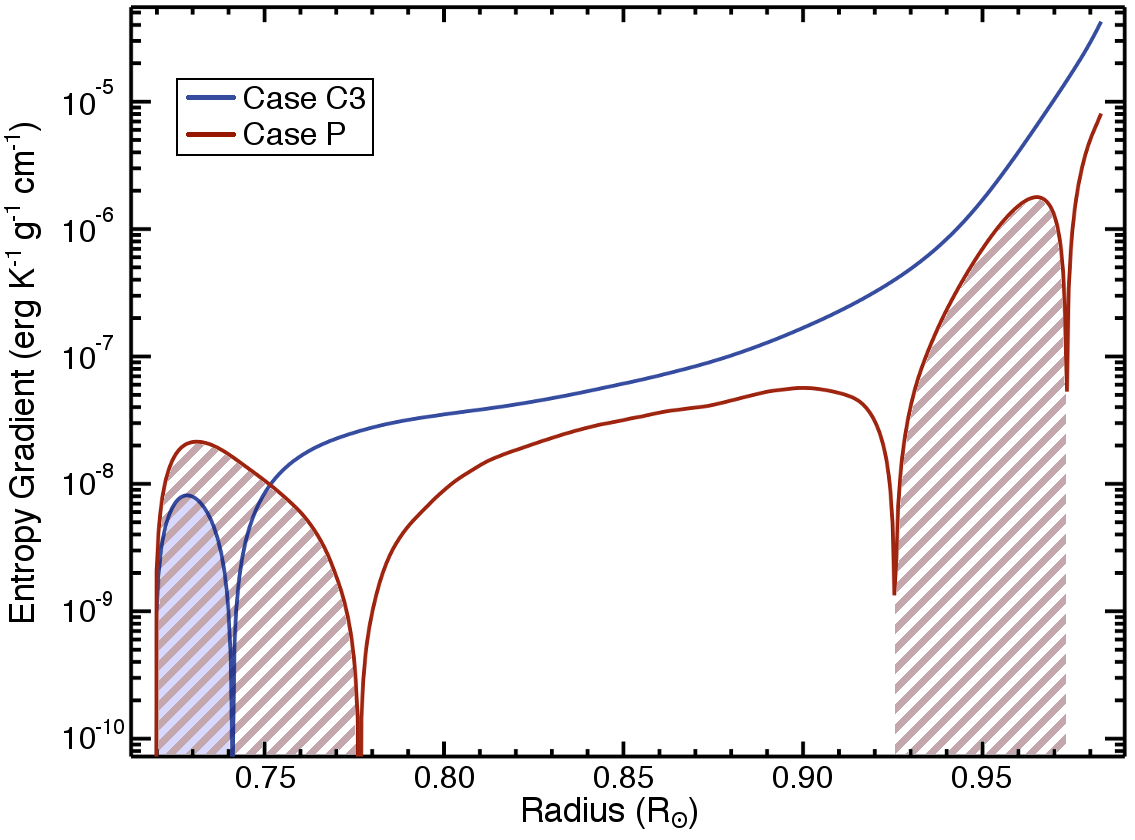}
\caption{Absolute value of the mean entropy gradient established by the convective flows in cases C3 (blue line) and P (red line). Regions where the stratification is subadiabatic are indicated by the area under the curves with hatched lines in corresponding colors. Overall case P exhibits much weaker super-adiabatic gradients than case C3, including substantial portions of the domain which are subadiabatic. \label{fig:dsdr}%
}
\end{center}
\end{figure}

It is particularly interesting to examine the behavior of plumes as they enter through the upper boundary in case P. The enthalpy flux experiences a sharp decline in the upper $0.01 R_\odot$ of the domain while the kinetic energy transport does not. However, both the enthalpy and kinetic energy contributions dramatically increase in magnitude over the upper quarter of the domain, peaking near $0.91 R_\odot$. This is a consequence of the imposed velocity and thermal plume structures. One clear short-coming of these plumes is that they are initially very smooth when real plumes would likely have well-developed secondary instabilities driving additional turbulence due to shear. In the future the use of plumes based on near-surface convection simulations may permit plume profiles which effectively include these sorts of self-consistent features.

In the bulk of the convective domain substantial changes in convective and diffusive fluxes are seen when comparing cases C3 and P. The enthalpy and kinetic energy transport both increase in magnitude in case P, mostly to offset each other but also to compensate for the drop in the transport by thermal diffusion. Thus the net convective energy transport is slightly higher in the bulk of the domain and significantly higher above $0.90 R_\odot$. At $0.97 R_\odot$ only a fifth of the solar luminosity is carried by convection in case C3, while in case P at the same depth the entire solar luminosity is carried by the sum of the enthalpy and kinetic energy fluxes.

Figure~\ref{fig:dsdr} shows the absolute value of the specific entropy gradient in cases C3 and P. These data have been averaged over both spherical surfaces and in time. Because the absolute value has been plotted, changes in sign are indicated by the abrupt dips seen, for example, at $0.74 R_\odot$ for case C3. In case C3 only a small region at the base of the convection zone is subadiabatic. In case P the stratification is superadiabatic above $0.973 R_\odot$, then subadabatic until $0.926 R_\odot$, then very weakly superadiabtic until $0.777 R_\odot$, and finally subabatic again to the base of the simulation.  Thus the bulk of the domain in case P contributes almost nothing to the convective driving.

\begin{figure}[t]
\begin{center}
\includegraphics[width=0.9\columnwidth]{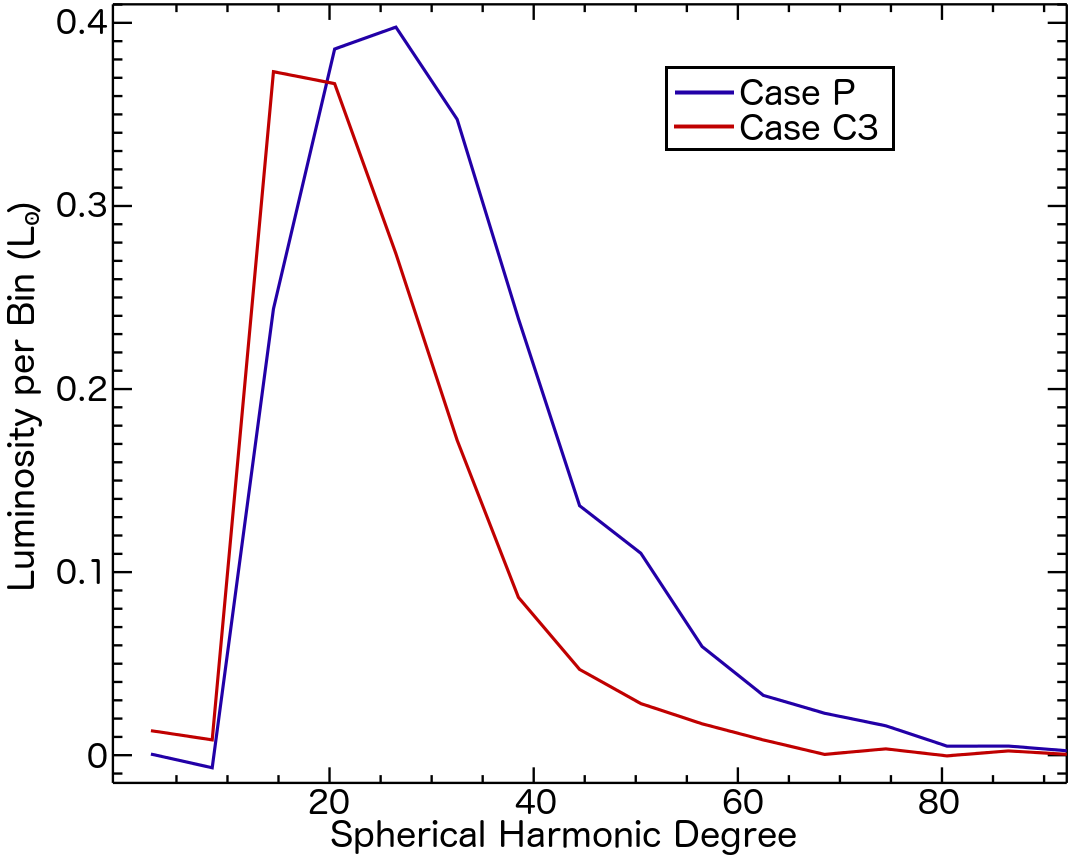}
\caption{Enthalpy flux spectra $\breve{\varepsilon}$ for cases C3 and P at mid-convection zone. Spectra have been binned with six modes per bin to reduce noise. Case P shows a clear shift in the peak of the enthalpy flux towards higher spherical harmonic degrees, indicating that smaller-scale convective structures dominate the convective energy transport. For visual clarity we have omitted contributions from spherical harmonics with $\ell > 90$ as their magnitudes, both individually and collectively, are small. \label{fig:EnSpec}%
}
\end{center}
\end{figure}

\begin{figure*}[t]
\begin{center}
\includegraphics[width=0.9\textwidth]{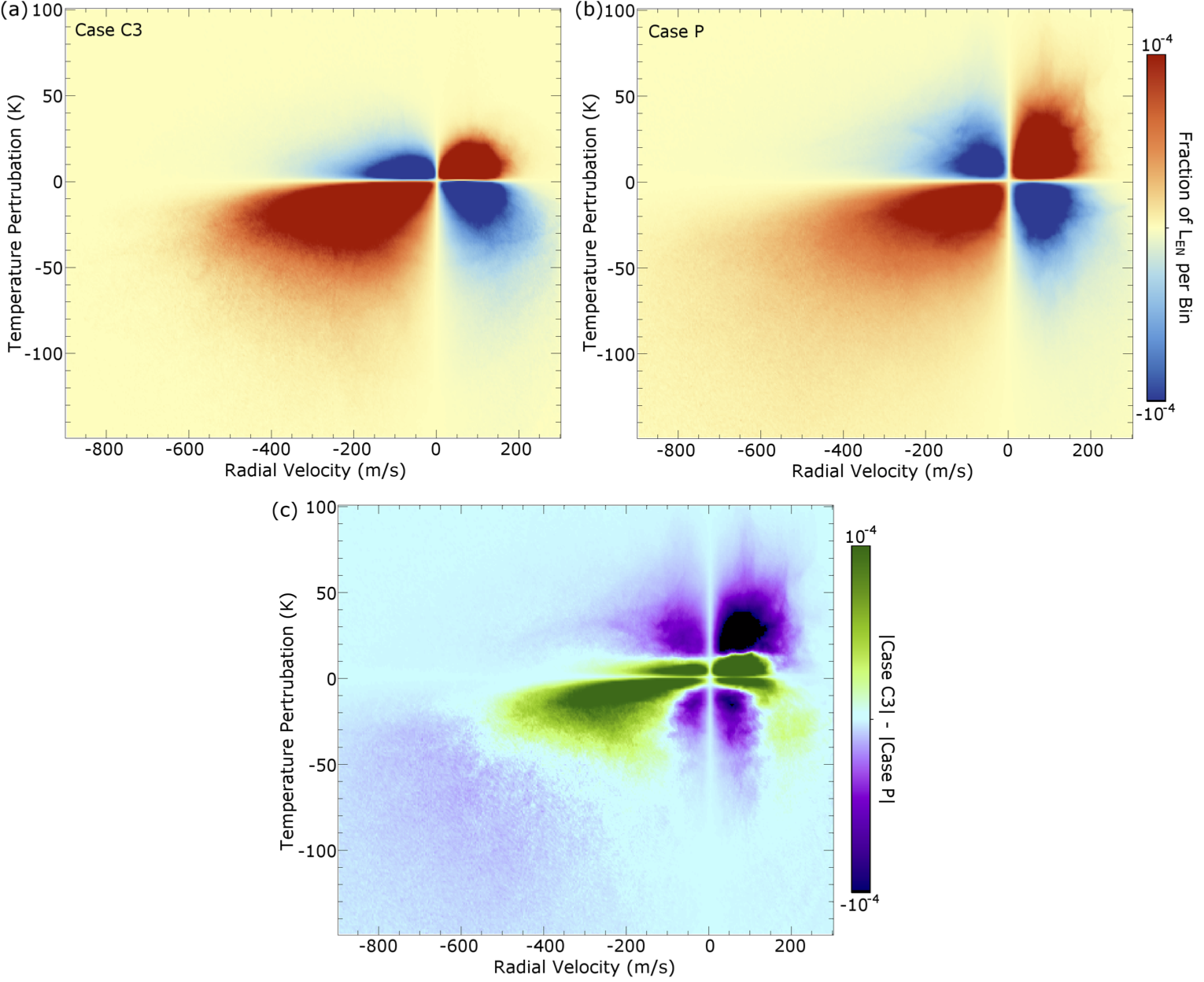}
\caption{Joint PDFs of radial velocity $u_r$ and temperature perturbation $T$ wighted to show the relative contribution of a given bin to the total enthalpy flux at mid-convections zone. Plotted are the joint probability distribution functions for enthalpy flux weighted by their enthalpy flux contribution for (a)~case C3 and (b)~case P. Blue tones for a given bin indicate inward enthalpy transport while red tones indicate outward transport. Quadrants I and III are by construction positive while quadrants II and IV are negative. (c)~The absolute value of (a) minus the absolute value of (b), providing a sense of the differences between the two figures. Bins where case C3 showed greater contribution to the net enthalpy flux are shown in green, while bins with a larger contribution from case P are given in purple. Case P shows more of the enthalpy flux contribution at more extreme flows both in the expected sense of colder, faster downflows and warmer, faster upflows, as well as more negative contributions from warm, fast downflows. \label{fig:PDFs}%
}
\end{center}
\end{figure*}

It is interesting to consider this result within the context of nonlocal convection models.  \citet{sprui97} argued that the structure and heat transport of deep convection may be dominated by small-scale plumes that form in the photospheric boundary layer and span the entire convection zone. If this were the case, then radiative heating of the broader, slower upflows could give rise to a subadiabatic stratification throughout most of the convective envelope, as noted by \citet{rempe05}.  Using a mean-field model, Rempel went on to show that such a subadiabatic stratification could promote baroclinic torques that sustain the conical solar differential profile against meridional flows that would otherwise establish a cylindrical (Taylor-Proudman) profile. \cite{Brandenburg2016} has formulated a mixing-length theory for this ``entropy rain'' convection and shown that it can be effectively employed in realistic solar structure models. \citet{hanas16} also advocated for this nonlocal, plume-dominated picture as a possible resolution of the `convection conundrum' discussed in \S\ref{sec:deep}.  This hypothesis is supported by the recent high-resolution, highly-stratified simulations of \citet{cosse16}.  When they confined the strong superadiabatic stratification to the surface layers, \citet{cosse16} found a substantial reduction in the large-scale convective power.  However, these simulations were Cartesian, non-rotating, and not fully equilibrated.  So, if the deep solar convection zone is indeed dominated by plumes, it remains uncertain how the convection could establish the solar differential rotation profile and carry out the solar luminosity.  Our results contribute to this evolving narrative by showing that plumes driven in the surface layers can self-organize into large-scale giant cells (\S\ref{sec:giants}) and establish a slightly subadiabatic stratification through much of the convection zone.

We have already demonstrated some significant differences in the net transport properties of the convection in case P, but we would like to further emphasize that there are major changes in the scales contributing to the convective energy transport. To see this we can look at the enthalpy flux spectra $\breve{\varepsilon}$ as a function of spherical harmonic degree $\ell$. Because the enthalpy flux is a binary product of radial velocity and temperature perturbations products of the same spherical harmonic degree and order can yield a net radial contribution. Thus we can compute the contribution of each spherical harmonic mode to the net radial transport as
\begin{equation}
\breve{\varepsilon}^2 \left( r, \ell \right) = \frac{1}{\Delta t} \int_{t_s}^{t_e} \sum_{m=0}^{\ell} \left| \breve{v}_r \left( r, \ell, m, t \right) \breve{T}^* \left( r, \ell, m, t \right) \right|^2 dt ,
\end{equation}
where the breves represent spherical harmonic coefficients of the associated physical quantities. 

Figure~\ref{fig:EnSpec} shows the enthalpy flux spectra $\breve{\varepsilon}$ at mid-convection zone for cases C3 and P when averaged over the same time interval in Figure~\ref{fig:EnSpec}. We have re-scaled $\breve{\varepsilon}$ to be in units of $L_\odot$ per mode. It is immediately clear that there has been a fundamental shift in the scales which dominate the enthalpy transport. In case P both the peak and the high-$\ell$ side of the power-containing modes are shifted to smaller scales. This indicates a significant change in the nature of the convection realized through the bulk of the convective layer in case P. For reference, the plumes applied on the outer boundary have their power and their enthalpy flux spectra peaked at $\ell \approx 60$, so the enthalpy flux at mid-convection zone is not simply due to coherent ``super-plumes'' which extend through the domain. Case P also shows a substantially greater contribution from modes with $\ell \gtrsim 60$, hinting at a substantial role for extreme, small-scale plumes which are not seen in similar closed boundary cases. Instead of a replacement of giant cell convection in case P, we are seeing a modification as smaller scales play an enhanced role and larger-scale correlations see their contributions diminished in the net radial transport of enthalpy. \\

\subsection{Roles of Upflows, Strong Plumes, and the Downflow Network}

Turbulent convection has generally defied reductionist descriptions, but for our purposes it is sometimes useful to think of giant cell convection as consisting principally of three components: a network of downflows lanes, the upflows bounded by the downflow lanes, and the strong plume-like downflows which are often found at the interstices of downflow lanes. Upflows are any location with $u_r > 0$, downflow lanes are regions where $u_r$ is between zero and roughly the negative RMS velocity, and the strong plumes are regions where $u_r$ is less than the negative RMS velocity. While inexact, these broad classes of flows are useful when comparing the convective flows in the bulk of the convective layer between cases C3 and P. 

\begin{figure}[ht]
\begin{center}
\includegraphics[width=0.95\columnwidth]{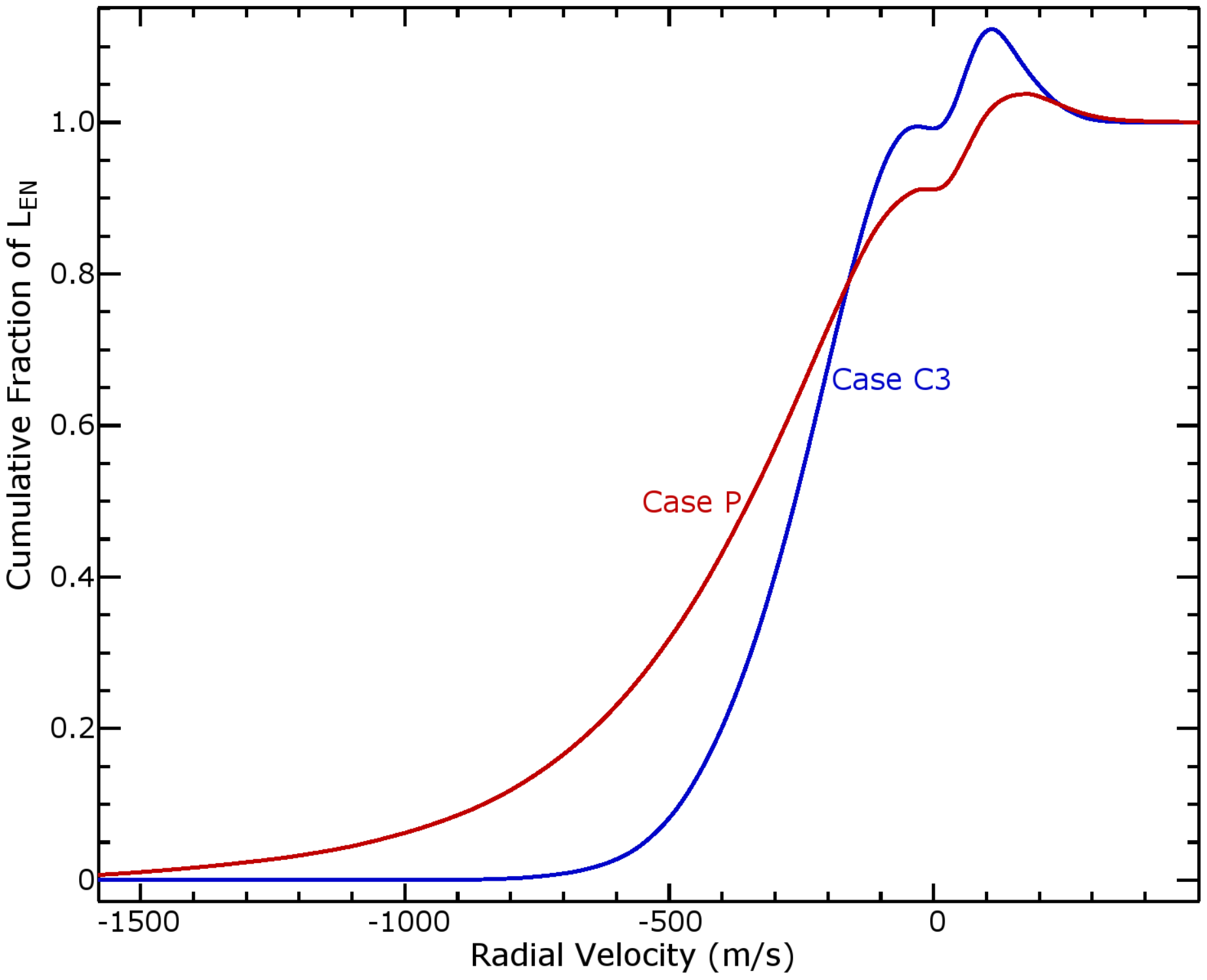}
\caption{Cumulative distribution function for the total enthalpy flux with respect to radial velocity seen at mid-convection zone for cases C3 and P. Case P shows almost half of its total enthalpy flux coming from downflows with speeds greater than 400 m/s, while case C3 shows about 15\% of its total enthalpy flux for downflows below that speed. Both cases show a net negative contribution from upflows above 100 m/s for case C3 and above 150 m/s for case P, though case C3 shows a much larger net effect. \label{fig:CDF_Vr}
}
\end{center}
\end{figure}

We have constructed joint probability distribution functions (JPDFs) of the radial enthalpy flux at mid-convection zone. These JPDFs were averaged in time over the same intervals used above. Figure~\ref{fig:PDFs}(a-b) show JPDFs of enthalpy flux at mid-convection zone scaled by the fraction of the total enthalpy flux transport per bin for cases C3 and P. Thus the color is proportional to the temperature perturbation and radial velocity of each bin times the fraction of the spherical surface covered by that bin (see Equation~\ref{eq:enthalpy}). By definition bins in the first and third quadrants are positive while bins in the second and fourth quadrants are negative. Bins near $T = 0$ or $u_r = 0$ are by definition zero. In both cases the dominant contribution to the total enthalpy flux is dominated by cold downflows which transport $2.24 L_\odot$ in case P and $1.81 L_\odot$ in case C3. The next largest contribution comes from hot upflows which transport $0.88 L_\odot$ in case P and $0.60 L_\odot$ in case C3. The largest negative contribution to the transport comes from cold upflows with $-0.72 L_\odot$ in case P and $-0.58 L_\odot$ in case C3. The smallest net contribution comes from warm downflows which yield $-0.51 L_\odot$ in case P and $-0.33 L_\odot$ in case C3. Cold upflows and warm downflows are generally related to either the entrainment or thermal diffusion by thermally buoyant flows.

While both cases see the same general trends, some clear differences emerge when we compare them in detail. Figure~\ref{fig:PDFs}(c) shows the difference of the absolute values of the JPDFs for cases C3 and P. Thus regions in green show bins where case C3 has a more dominant contribution to enthalpy transport than case P and vice versa for purple regions. Case P shows far more enthalpy transport by convective parcels with large temperature perturbations, particularly for negative $T$. There is a roughly horizontal line at about $+15$ K which demarcates case C3's dominance from case P's for almost all radial velocities. This seems to indicate a change in the entrainment properties of convective downflows and possibly in the driving of convective upflows which in case C3 occur through bulk convective excitation by the superadiabatic gradient, while in case P upflows occur essentially only as a response to the plume driving. The excess in case P for transport by upflows warmer than $+15$ K is particularly striking.  On the cold side of the distribution, we see case C3 dominating for cold downflows except for extreme events, which are virtually non-existent in case C3, and for some low-velocity extreme-temperature plumes. Warm downflows are dominated by case C3 for small temperature perturbations and for radial velocities above about 150 m/s, while case P again dominates for small-velocity, large-temperature events. 

\begin{figure}[t]
\begin{center}
\includegraphics[width=0.95\columnwidth]{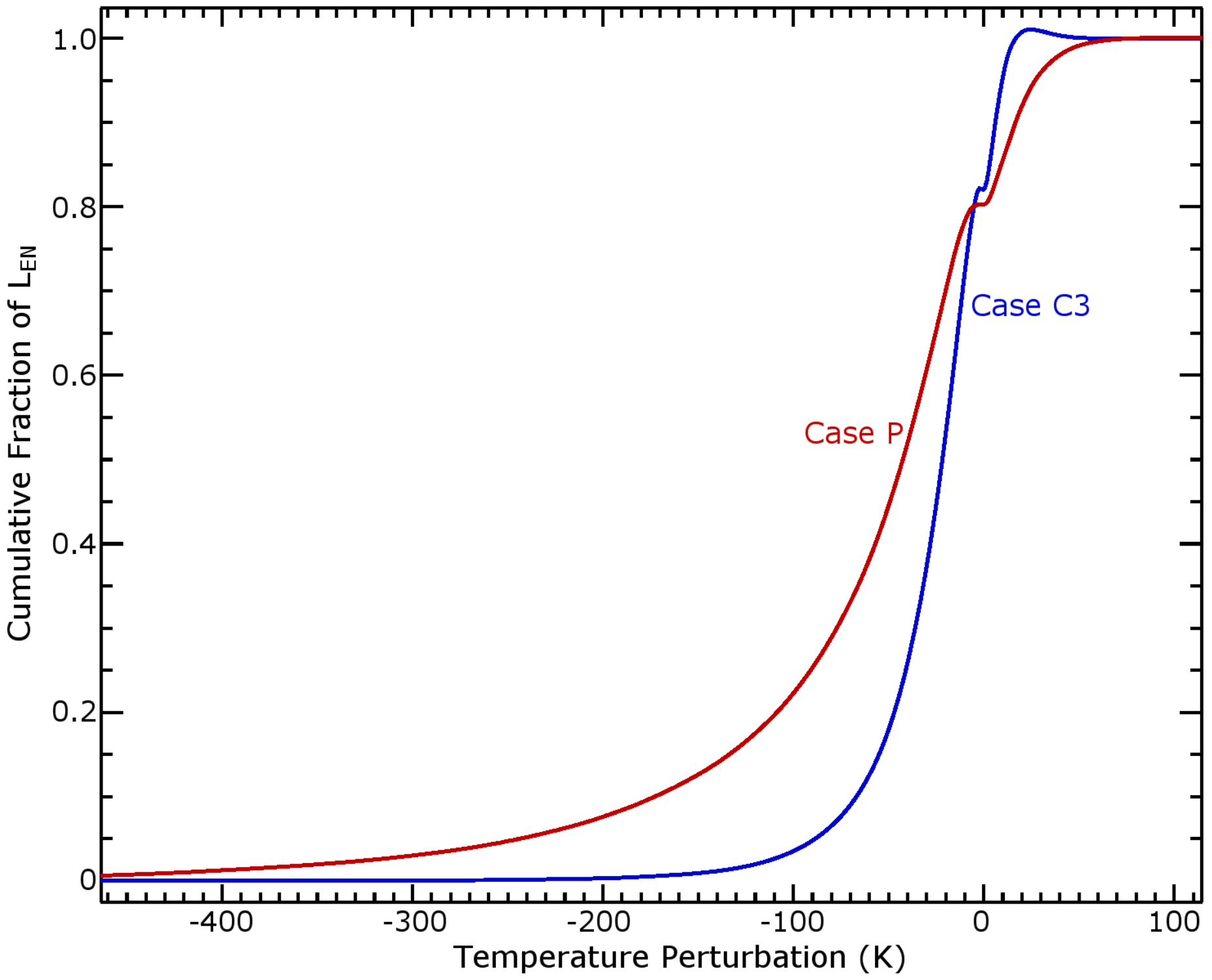}
\caption{Cumulative distribution function for the total enthalpy flux at mid-convection zone with respect to temperature perturbation for cases C3 and P. As in Fig.~\ref{fig:CDF_Vr}, case P shows a much larger contribution from extremely cold regions. Case P exhibits a monotonic function while for case C3 bins above 20 K show a net negative contribution. \label{fig:CDF_T}%
}
\end{center}
\end{figure}

\begin{figure*}[t]
\begin{center}
\includegraphics[width=0.85\textwidth]{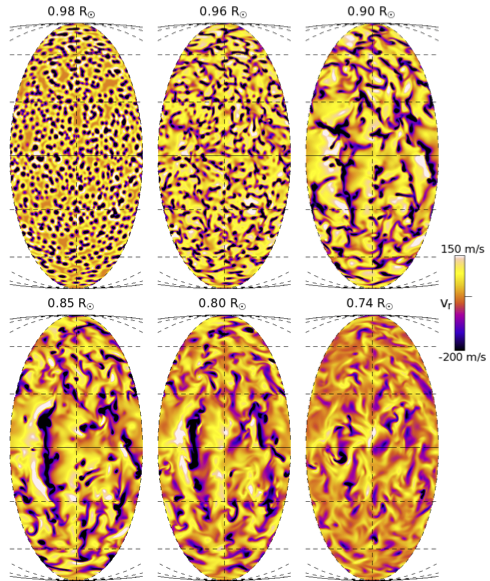}
\caption{Snapshots of radial velocities for case P on spherical shells at the indicated radii. Downflows begin at the boundary as small, randomly placed plumes and quickly begin to coalesce into larger structures. By $0.90 R_\odot$ significant north-south alignment can begin to be seen at low latitudes. By mid-convection zone a coherent giant cell can be seen near the equator. \label{fig:VrSlices}%
}
\end{center}
\end{figure*}

Another way to examine the statistical differences seen in the convection between cases C3 and P is to again look at the enthalpy transport but instead look at a cumulative distribution function (CDF) -- the fraction of the total enthalpy luminosity transported through all regions of a given surface with some parameter less than a given value. Figures~\ref{fig:CDF_Vr} and \ref{fig:CDF_T} show the CDFs for enthalpy flux with radial velocity and temperature, respectively, again at mid-convection zone. Figure~\ref{fig:CDF_Vr} demonstrates two major changes from the stratification-driven convection in case C3 and the plume-driven convection in case P. First, the plume-driven convection shows dramatically increased enthalpy transport for ultra-fast downflows with radial velocities below -500 m/s. These extreme downflows contribute about 35\% of the total enthalpy transport in case P, while the same flow speeds contribute less than 10\% of the total enthalpy flux in case C3. Second, while the CDF for case C3 shows that the  approximately  total enthalpy transport through the mid-convective zone. Thus the upflows in case C3 have essentially zero net contribution to the enthalpy transport. In contrast the upflows in case P produce a net transport of a little more than 10\% of the total enthalpy flux. Figure~\ref{fig:CDF_T} shows similar comparative trends in the CDF of total enthalpy flux with respect to temperature perturbations. Ultra-cold regions with $T < -100$ K play an important role in plume-driven convection and essentially zero role in stratification-driven convection, while warm regions play a larger net positive role in case P compared to case C3.

Taken together, the analysis of the convective transport of enthalpy presented here supports the conclusion that modifying the upper-boundary condition and hence the entropy gradient through the convective layer produces clear differences in the resulting convective structures at mid-convection zone. Cold and inward convective plumes become statistically more extreme in our plume-driven model. Both extreme downflows and all upflows shift to have greater net positive contributions to the radial transport of the solar luminosity. In general the convection becomes less reliant on the downflow network to transport the solar luminosity through the bulk of the convective layer while strong plumes and upflows play larger roles. \\

\section{Coalescence of Plumes into Giant Cells} \label{sec:giants}

After an examination of the differences in the spectral and statistical properties of the convective energy transport between stratification-driven and plume-driven convection in the bulk of the convection zone, we now turn to a surprising similarity. Despite the change in driving, the plume-driven convection in case P still yields convective structures similar to the giant cells which have long been seen in global convective models \cite[e.g.,][]{Brun2002, Miesch2005, Ghizaru2010, Nelson2013a, Featherstone2015}. The appearance of giant cells in deep convective models has long been explained as a global response to bulk driving by a super-adiabatic gradient, however another possibility for their emergence is the self-organization of near-surface plumes as a response to the density stratification \citep{Nordlund_Stein_Asplund_2009}.


We do indeed see such self-organization, as illustrated in Figure~\ref{fig:VrSlices}.  This shows radial velocity patterns for case P at six depths spanning the simulated domain using a constant color scale for ease of comparison between depths. All six snapshots are taken at the same instant. At $0.98 R_\odot$ we see the plumes a single grid point in from the outer boundary condition. Moving inward, by $0.96 R_\odot$ the plumes begin to merge and interact. At $0.90 R_\odot$ the small-scale plumes have given way to much larger complexes of downflows which are beginning to resemble lanes and have hints of the cellular pattern seen in closed-boundary simulations. At mid-convection zone clear north-south alignment can be seen at low latitudes. This persists through $0.80 R_\odot$ until finally the closed boundary at the base of the domain wipes out these patterns by $0.74 R_\odot$. 

\begin{figure*}[t]
\begin{center}
\includegraphics[width=0.9\textwidth]{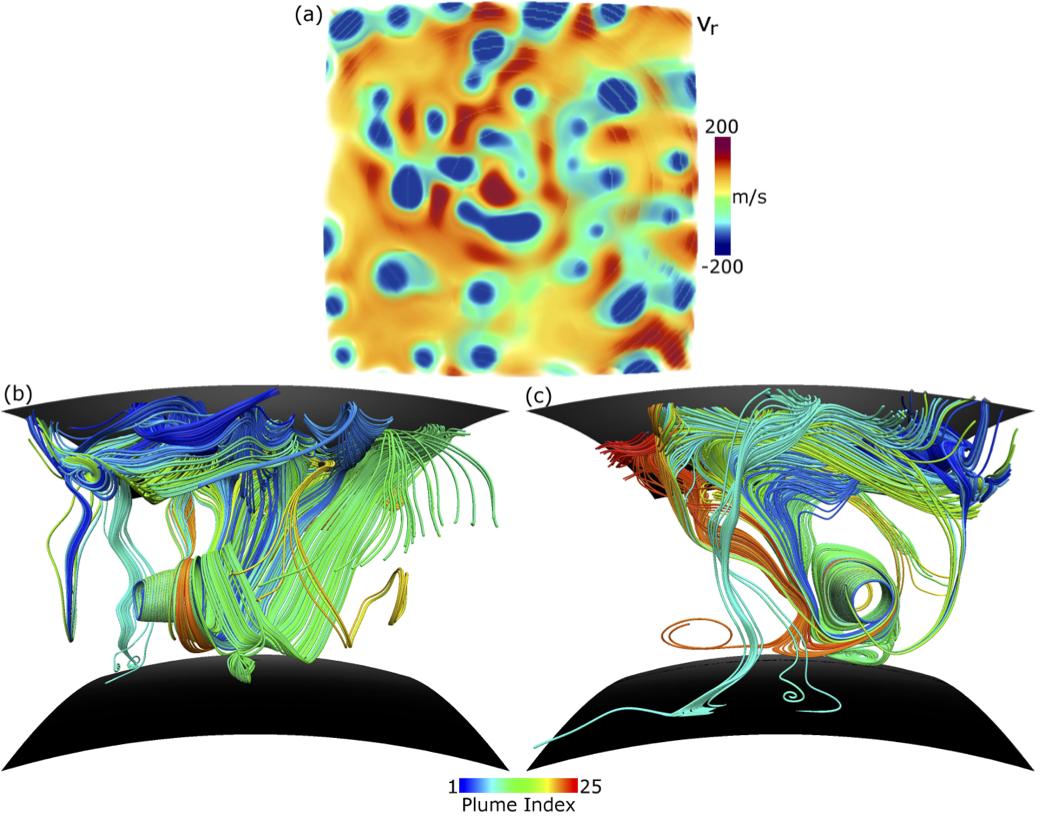}
\caption{(a)~Snapshot of radial velocity on the outer boundary of case P over a $30^\circ$ by $30^\circ$ patch centered on the equator with north pointing up and east pointing to the right. This patch contains 25 imposed plumes including three in the ramp-up phase and 5 in the decaying phase of their lifetimes. Plumes are numbered from 1 to 25 based on their central longitude from west to east. The downflow portion of each plume is seeded with 50 streamlines randomly initiated at between $0.973$ and $0.983 R_\odot$. Streamlines (lines tangent to the velocity field at a single instant) are then integrated downward into the simulation domain. (b)~3D rendering of the streamlines initiated in the boundary plumes colored by the assigned plume number. Perspective is looking west. The upper and lower boundaries of the simulation domain have been added as black surfaces for visual clarity. (c)~Same as (b) but now looking north. Almost all of the 25 plumes coalesce into a large rotationally-aligned downflow lane which is part of a strong giant cell in the bottom half of the domain. \label{fig:streamlines}%
}
\end{center}
\end{figure*}

The convective patterns seen in Figure~\ref{fig:VrSlices} are suggestive of self-organization into structures which are both visibly and quantitatively similar to giant cells seen in closed boundary simulations. At $0.90 R_\odot$ where there is a relatively strong subadiabatic gradient, there are clear indications of downflow lanes cellular patterns. At $0.85 R_\odot$ a strong, rotationally aligned banana cell spans over $70^\circ$ in latitude around the equator. These are the hallmarks of giant cell convection seen in models with uniform driving at the boundaries.

In some ways the emergence of convective motions dominated by scales larger than those imposed by our plume-driven model is not unexpected. The increase of the density scale height and the effects of a convergent geometry tend to shift convective structures to larger scales. What is striking is the coalescence of the individual plumes into downflow lanes. The plumes choose to form horizontally anisotropic sheets rather than simply larger-scale plumes. Even at high latitudes where rotational influences are insufficient to drive rotational alignment there are still clear indications of cellular patterns.



Perhaps even more striking evidence of this self-organization can be seen by tracing instantaneous streamlines through our simulation to see how the imposed plumes link to the giant cell convective patterns seen in Figure~\ref{fig:VrSlices}. Figure~\ref{fig:streamlines} shows an example of this connectivity between boundary-forced plumes and deep convective cells. Figure~\ref{fig:streamlines}a shows the radial velocity imposed by our plume boundary condition over a 30$^\circ$ by $30^\circ$ patch of case P. At that time there were 25 imposed plumes in the patch, most of which overlapped with at least one other imposed plume. Each plume's core downflow region was seeded with 50 streamlines which were colored based on the position of the plume. Thus all streamlines from plume 1 (located in the bottom left corner of Figure~\ref{fig:streamlines}a ) were colored blue. The streamlines were than traced down into the domain. Figure~\ref{fig:streamlines}b and c show two views of the resulting streamlines looking west and north at the same instant. The streamlines in the lower half of the convective layer highlight the presence of one side of a rotationally-aligned banana cell which is here seen as the vortex-like structure in panels b and c.

\begin{figure*}
\begin{center}
\includegraphics[width=\linewidth]{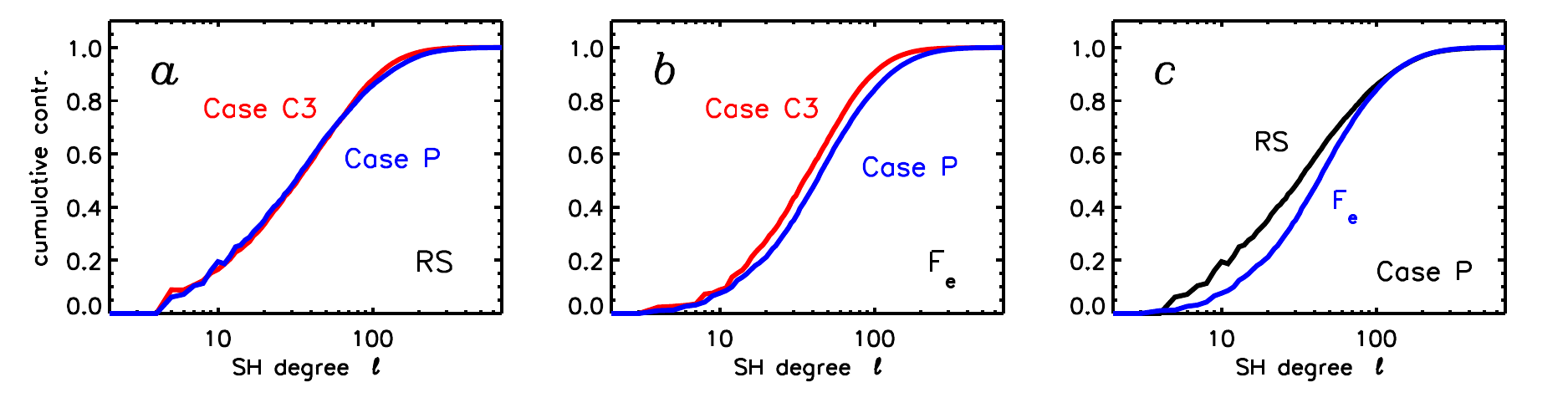}
\caption{Cumulative spectra computed by applying a spectral filter as described in the text. (\textit{a}) Latitudinal angular momentum transport by the convective Reynolds stress, $C(\ell)$ for Case C3 (solid line) and case P (dashed line). (\textit{b}) As in (\textit{a}) but for the convective enthalpy flux $\left< v_r^\prime T^\prime \right>$.  (\textit{c}) The Reynolds stress spectrum $C_\theta(\ell)$ (solid line) is plotted together with the enthalpy flux spectrum (dashed line) for case P.  All curves are computed in the mid convection zone ($r = 0.85 R_\odot$) and are averaged over time.\label{fig:rs_spectrum}
}
\end{center}
\end{figure*}

Of the 1250 streamlines traced from random seeds in the cores of the imposed plume downflows, 871 connect to a depth of $0.75 R_\odot$. A remarkable 736, including at least one streamline from 23 of 25 plumes, connect to the large banana cell in the bulk of the convective layer. This demonstrates a high degree of linkage between the imposed plumes from the upper boundary. It is clear from case P that plume-driven convection can readily coalesce into large-scale convective cells and that large-scale convective cells can readily connect to small-scale plumes, providing an interesting and potentially more realistic means to drive global solar convection models which can also connect with observational constraints.

Giant cell convection had often been modeled as the non-linear saturation of global-scale convective instabilities arising due to a super-adiabtic gradient through the bulk of the convection zone. Recent claims of detection of giant cell convective motions have reinforced this view as their measurements have been consistent with the predicted horizontal scale from boundary-driven convective models \cite{Hathaway_Upton_Colegrove_2013, McIntosh_Wang_Leamon_Scherrer_2014, Greer2015}, although the amplitude of these motions appears to be much less than predicted \citep{lord14,hanas16}. 

In any case, our plume boundary simulations suggest that giant cell convective patterns can arise in a rotating convection zone without strong super-adiabatic gradients.  It remains to be seen whether this result is consistent with the results of \citet{cosse16}.  They argued that power at large scales was reduced when the bulk of the convection zone was adiabiatically stratified.  However, these were non-rotating simulations that were not in equilibrium in the sense that they did not carry the full solar luminosity through the entire convection zone.   Furthermore, as argued by \citet{Featherstone2016}, giant cells may not be so giant after all.  If the Rossby number of the deep convection zone is very low, the peak of the power spectrum might shift to scales that are no bigger than supergranulation, $\ell \sim 120$.  However, what matters is that, unlike supergranulation, these giants cells are highly anisotropic, with a preferential alignment parallel to the rotation axis.  This is what is needed to produce the convective Reynolds stresses necessary to sustain the solar differential rotation.  We have demonstrated that such rotationally-influenced convective structures (banana cells) can coexist with and even be driven by a network of smaller-scale, isotropic plumes that originate in the surface layers.

It also remains to be seen how our work fits with the work of \citet{hotta14,hotta15}.  In high-resolution, highly-stratified, non-rotating convection experiments, \citet{hotta14} found that the power spectrum and heat transport in the mid convection zone was insensitive to the presence of small-scale surface convection.  Simulations with an upper boundary of 0.99$R_\odot$ gave similar results to simulations with an upper boundary at 0.96$R_\odot$.  This appears to contradict our result that the dominance of small-scale plumes near the surface leads to a change in the nature of the heat transport throughout the shell.  However, \citet{hotta14} did not present the entropy gradients in these cases so it is difficult to judge how close the deep stratification was to adiabatic and how much this contributed to the buoyancy driving.  They also did not present spectra of the enthalpy flux so it is difficult to compare their results directly with ours.  However, their subsequent rotating simulations did exhibit a transition from small-scale, nearly isotropic, plume-dominated convection near the surface to more rotationally-aligned giant cells in the mid convection zone, as we have found here \citep{hotta15}.

As argued, e.g.\ by Miesch (2005), the axial alignment of rotationally-constrained giant cells (i.e.\ banana cells with a north-south orientation at low and mid latitudes) is essential to produce solar-like differential rotation in global convection simulations (see Sec.\ 1.2).  The Coriolis-induced Reynolds stress associated with such alignment give rise to convective angular momentum transport toward the equator, which sustains the equatorward $\Omega$ gradient.

Though the solar-like differential rotation in Case P is sustained in large part by the external forcing, as discussed in Sec.\ 2.3, the presence of banana cells is encouraging.  In particular, it raises the possibility that relatively weak, large-scale, rotationally constrained (Ro $\ll 1$) convective motions (namely banana cells) can provide that angular momentum transport that sustains the differential rotation while smaller-scale motions (namely plumes) account for the convective heat transport that carries the solar luminosity.  This is one potential way out of the convection conundrum discussed in Section 1.

Does this scale separation between convective angular momentum transport and heat transport occur in Case P?  This question is addressed by Fig.~\ref{fig:rs_spectrum}.  These are cumulative spectra similar to the spectra in Figs.~7 and 12 but computed in a different way.  For example, Fig.~7 was computed as $\sum_m \breve{v}_r(r,\ell,m,t) \breve{T}(r,\ell, m,t)$, where breves again indicate the expansion coefficients obtained from spherical harmonic transformations.  Such a procedure for the latitudinal Reynolds stress would yield only residual noise due to the symmetry properties of $v_\theta$ and $v_\phi$.

So, we instead compute the Reynolds stress spectra by applying a low-pass spectral filter to each of $v_\theta(r,\theta,\phi,t)$ and $v_\phi(r,\theta,\phi,t)$ that removes all power in spherical harmonic modes with $\ell > \ell_f$, where $\ell_f$ is a threshold wavenumber that is varied from 0 to $\ell_{max}$.  After applying the low-pass filter, we then compute the latitudinal component of the Reynolds stress as $R_\theta(r,\theta,t;\ell_f) = \left<v_\theta^\prime v_\phi^\prime\right>$ where primes indicate departures from the mean flow, e.g.\ $v_\theta^\prime = v_\theta - \left<v_\theta\right>$.  We then define the cumulative spectrum as
\begin{equation}\label{eq:Cell}
C_\theta(\ell_f) = \frac{1}{t_2-t_1} \int_{t_1}^{t_2} \left( \frac{\int_0^\pi {\vert R_\theta(r_0,\theta,t;\ell_f)} \vert \sin\theta d\theta}{\int_0^\pi {\vert R_\theta(r_0,\theta,t;\ell_{max})} \vert \sin\theta d\theta} \right)  dt ~,
\end{equation}
where $r_0 = 0.85 R_\odot$ is a chosen radial level in the mid convection zone and the time interval $t_1 \rightarrow t_2$ spans either 18 days (Case P) or 36 days (Case C3).  So, $C(\ell_f)$ ranges from zero for $\ell_f = 0$ to unity for $\ell_f = \ell_{max}$ and the rate at which it rises reflects contributions to the transport from a particular scale.  The function $C_\theta(\ell_f)$ is shown in Fig.\ \ref{fig:rs_spectrum}\textit{a} for Cases C3 (solid line) and P (dashed line).  Note that we have dropped the subscript of $\ell_f$ from the abscissa label.

Figure \ref{fig:rs_spectrum}\textit{a} indicates that the scales responsible for convective angular momentum transport are the same in cases C3 and P.  We saw above with Fig.\ 7 that this is not the case for the convective heat flux, which is shifted toward smaller scales in Case P.  The cumulative spectrum for the enthalpy flux, computed as for $C_\theta(\ell_f)$ as above, shows a similar shift toward smaller scales.  This is demonstrated in Fig.\ \ref{fig:rs_spectrum}\textit{b}.  The shift is not dramatic, but it is significant.  Furthermore, when the Reynolds stress spectrum $C_\theta(\ell_f)$ in case P is compared with the corresponding enthalpy flux spectrum, it is clear that the latter occurs at significantly higher wavenumber. While the similarities in Reynolds stress spectrum between cases C3 and P provide confidence that the dynamics seen in case P are not dominated by the corrective torque applied at the boundary, we must nevertheless consider that our choice of parameters for the corrective torque may play some role here.

Though the effect is not nearly enough to solve the convection conundrum, we have for the first time demonstrated scale separation between angular momentum transport and heat transport in a global solar/stellar convection simulation.  At the extreme parameter regimes characteristic of actual stars, we might expect this scale separation to be much more pronounced.

\section{Discussion}

In this paper we have presented a novel stochastic boundary condition which mimics aspects of near-surface convective flows descending into the bulk of the solar convection zone. A simulation with this plume boundary model has shown that giant cell convection need not arise from the classical Rayleigh-Bernard-like scenario where convective driving occurs through strong entropy gradients near domain boundaries. Instead we have applied small-scale plumes designed to mimic those from near-surface convection. This has led to significant changes in the resulting convection, however the plumes have also shown an ability to self-organize into cellular convective structures on scales large enough to be rotationally constrained.

The plume boundary condition leads to giant cell convection that differs from similar closed boundary simulations in several important ways. We have shown that the plume boundary condition greatly reduces the super-adiabatic gradient. Over significant portions of the simulation we even see a slightly sub-adiabatic gradient with the plume boundary condition \citep[see][]{Hotta2017, Kapyla2017,Korre2017,Bekki2017}. We also see a dramatic reduction in the diffusive boundary layer with the removal of the closed-boundary constraint. In its place we see a new type of boundary where the imposed small-scale plumes match onto the simulation-generated flows rising from deeper layers. Using an spectral analysis of the enthalpy flux, we have shown that the convective transport in our plume boundary simulation has shifted to smaller scales, indicating that the large-scale convective motions play less of role in this process. That conclusion was further confirmed by an examination of the distribution of the enthalpy flux over velocity and temperature perturbations in both closed and plume boundary simulations showing that the plume boundary simulation showed greatly enhanced transport from both upflows and extreme downflows.

Perhaps more striking than the differences between simulations was the result that small-scale plume driving still resulted in convection dominated by giant cells at mid-convective zone, including clear banana cells at low latitudes. The self-organization of plumes into sheets and cellular patterns, along with the connectivity in streamlines between giant cells and the imposed plumes demonstrates that giant cell convection can in fact occur under the paradigm of near-surface convective driving. This simulation is, to our knowledge, the first which both produces rotationally constrained convective structures (bannana cells) which are thought to be responsible for maintaining the solar differential rotation and is driven through nonlocal, small-scale plumes.  Furthermore, the scales responsible for transporting angular momentum by means of the Reynolds stress are essentially the same in our plume-driven simulation as in our corresponding diffusive simulation.

This point is worth emphasizing, that the heat flux is shifted toward smaller scales with the plume boundary condition but the angular momentum transport by the convective Reynolds stress is not.  We expect this scale separation to be more pronounced at the extreme parameter regime of the solar interior.  This may help resolve the convection conundrum by shifting the burden of heat transport to scales that are not resolved in current global convection simulations.  Large scales would then possess reduced velocity amplitudes and consequently the low Rossby number needed to establish a solar-like differential rotation. 

Our plume boundary condition is limited in its treatment of angular momentum conservation and the differential rotation achieved in case P is strongly related to the influence of both the flux of angular momentum through the outer boundary and the corrective torque applied to compensate for that flux. In spite of those limitations, it is encouraging that at least for some parameter choices the plume-boundary model can maintain a solar-like differential rotation profile at levels of turbulence where closed-boundary models do not. Future work will seek to understand what impacts various additional formulations of the corrective torque and the applied boundary condition may have on these models.

Despite the inherent conceptual and numerical challenges of a stochastic, semi-open boundary condition in a simulation that is both anelastic and uses pseudospectral methods, the success of this model in achieving weak entropy gradients, and qualitatively similar giant cell convection far from the boundary suggest a new path forward in modeling global stellar convection. This work also highlights the continued need for substantial efforts to better link near-surface and deep convection models of the Sun.

\acknowledgements
The authors would like to thank Kyle Augstson, Benjamin Brown, Mark Rast, and Regner Trampedach for many useful insights and suggestions. NJN received support from the Nicholas C. Metropolis Post-Doctoral Fellowship as part of the Advanced Simulation Capabilities Initiative at Los Alamos National Laboratory. NAF recieved support from NASA grant NNX17AM01G, and NSF grants NSF-0949446 and NSF-1550901. Simulations were preformed on the Pleiades Cluster, operated by the NASA Science Mission 
Directorate High-End Computing program through award g26133.

\bibliography{full_article_arxiv.bbl}

\newpage
\begin{appendix}

\section{Implementation of the Plume Boundary Condition}

\begin{figure}[h]
\begin{center}
\includegraphics[width=0.7\columnwidth]{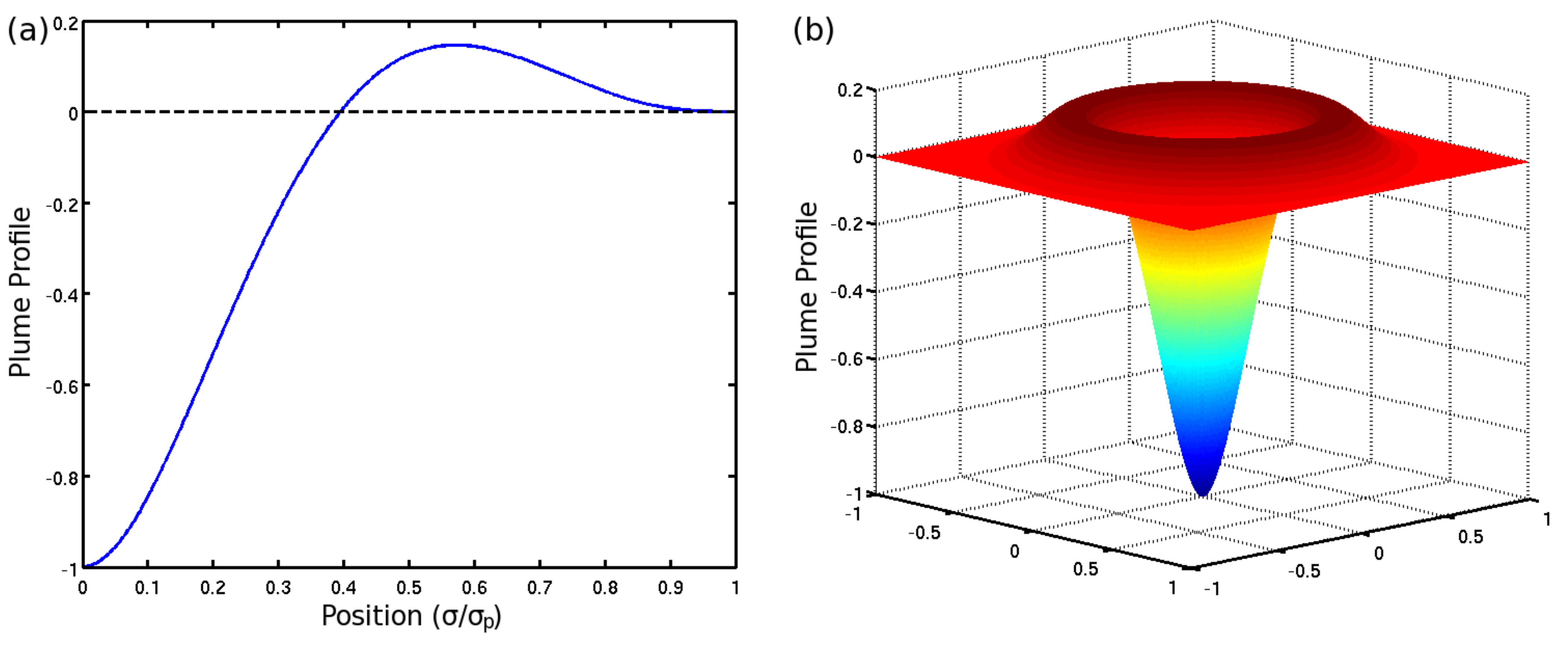}
\caption{(a)~Profile of the shape of the imposed plumes as a function of angular distance from the plume center $\sigma$ divided by the width of the plume $\delta$. (b)~3D surface rendering of the plume shape. The plumes consist of the strong negative core of downflowing low-entropy material surrounded by a weaker positive ring of higher entropy upflows. Surface integrals of this shape are identically zero by design.
  \label{fig:Profile}%
}
\end{center}
\end{figure}

The plume boundary conditions discussed in this paper consist of (a) prescribed, stochastic Dirichlet boundary conditions on the radial velocity and entropy fields, (b) a prescribed, stochastic Robin boundary condition relating the horizontal divergence of the velocity field to the radial velocity, and (c) a fixed Neumann boundary condition on the toroidal component of the mass flux. Component (c) corresponds to the toroidal part of the traditional stress-free boundary condition on horizontal velocities. Here we specify the mathematical structure of this set of boundary conditions. In addition, to ensure global angular momentum conservation we additionally apply a volumetric torque which is discussed in Section 2.2.

The boundary condition for radial momentum is
\begin{equation} 
u_r \left( R_o, \theta, \phi \right) = \mathcal{R} \left( \theta, \phi, t \right)
\label{eq:vr}
\end{equation} 
and the boundary condition for the entropy equation is likewise
\begin{equation} 
S \left( R_o, \theta, \phi \right) = \mathcal{S} \left( \theta, \phi, t \right) ,
\end{equation} 
where $R_o$ is the radius of the outer boundary and $\mathcal{R} \left( \theta, \phi, t \right)$ and $ \mathcal{S} \left( \theta, \phi, t \right)$ are functions of position on the spherical surface and time chosen to in at least some degree reflect properties of near-surface convective plumes. We then choose $\mathcal{R} \left( \theta, \phi, t \right)$ and $ \mathcal{S} \left( \theta, \phi, t \right)$ to be composed of $N_p$ small-scale plume structures, each with some velocity amplitude $\mathcal{V}$ and entropy amplitude $\mathcal{E}$. Thus we can write these fields as
\begin{equation}
\mathcal{R} \left( \theta, \phi, t \right) = \sum_{i=1}^{N_p} \mathcal{V}_i \, \mathcal{P}_i \left( \theta, \phi \right) \mathcal{T}(t)
\end{equation}
\begin{equation}
\mathcal{S} \left( \theta, \phi, t \right) = \sum_{i=1}^{N_p} \mathcal{E}_i \, \mathcal{P}_i \left( \theta, \phi \right) \mathcal{T}(t),
\end{equation}
 where the time dependence is encoded in $\mathcal{T}(t)$. The plume profiles are designed such that they have compact support (locally specified generating functions). This greatly reduces the computational cost of applying the plumes. Additionally, we choose $\mathcal{R} \left( \theta, \phi, t \right)$ and $ \mathcal{S} \left( \theta, \phi, t \right)$ such that
\begin{equation} 
\int_S \mathcal{R} \left( \theta, \phi, t \right) \sin \theta \, d \theta \, d \phi =  \int_S \mathcal{S} \left( \theta, \phi, t \right) \sin \theta \, d \theta \, d \phi = 0 ,
\end{equation} 
where the integrals are taken over the spherical surface. Thus our plumes carry no net mass flux and set the mean entropy on the boundary to zero. We employ a zero-flux plume profile $\mathcal{P}$ which is designed such that
\begin{equation} 
\int_S \mathcal{P} \sin \theta \, d \theta \, d \phi = 0.
\end{equation} 

We choose our locally mass-conserving plumes to follow the profile
\begin{equation}
\mathcal{P} \left( \psi \right) = 
   \begin{cases}
    -1 + 20 \psi^2 - 50 \psi^3 + 45 \psi^4 - 14 \psi^5 & \mathrm{ if } \; \psi \leq 1 \\
    0 & \mathrm{ if } \; \psi > 1
    \end{cases}
\end{equation}  
where $\psi$ is the angular distance from the center of the plume $\Delta \sigma$ divided by the width of the plume $\delta$. This polynomial representation is continuous in its value, and first and second derivatives at all points. Figure~\ref{fig:Profile}(a) shows the polynomial, and Figure~\ref{fig:Profile}(b) shows a 3D surface rendering of the plume profile. The profile has a large negative core of low-entropy downflow surrounded by a low-amplitude positive ring of higher entropy upflow. This mimics the shape of plume profiles seen in near-surface models \citep{Rast_1998}. We assign these plumes initially to uniform random locations on the outer boundary, accounting for the spherical geometry. Each plume is assigned an angular width $\delta$. 

We must then determine the angular distance from each grid point to each plume center in order to apply the piecewise function $\mathcal{P} \left( \Delta \sigma / \delta \right)$. Determining this distance over the surface of a sphere is a surprisingly difficult computational challenge as most methods suffer from significant numerical error at small seperations. We here use a numerical implementation of the classic Haversine formula designed to be accurate for small angular separations given by 
\begin{equation} 
\Delta \sigma \left( \theta, \phi, \theta_p, \phi_p \right) = 2 \sin^{-1} \left[ \sin^2 \left( \frac{ \Delta \theta }{ 2 } \right) + \cos \theta \, \cos \theta_p \sin^2 \left( \frac{ \Delta \phi }{ 2 } \right) \right] .
\end{equation} 
This implementation suffers errors for separations larger than $\pi / 2$ radians, but is highly accurate and computationally inexpensive for small angles.

We also choose to have our plume field vary in time. Each plume is assigned a lifetime $\tau$. The plume's amplitude is then modified by $\mathcal{T} \left( t, t_0, \tau \right)$, which is defined as a function of the time since the plume was initiated $t - t_0$ by
\begin{equation}
\mathcal{T} \left( t,  t_0, \tau \right) = 
   \begin{cases}
    \frac{ t - t_0 }{ 0.2 \tau } & \mathrm{ if } \: 0 \leq \frac{ t - t_0 }{ \tau } < 0.2 \\
    1 &  \mathrm{ if } \: 0.2 \leq \frac{ t - t_0 }{ \tau } < 0.8 \\
    1 - \frac{ t - t_0 - 0.8 \tau}{ 0.2 \tau } &  \mathrm{ if } \: 0.8 \leq \frac{ t - t_0 }{ \tau } < 1.0  \\
    \end{cases} .
    \label{eq:plume time}
\end{equation}  
This provides linear ramp-up and cool-down phases for the plumes, minimizing the spurious pressure perturbations which can plague these simulations. A plume expires when $t - t_0 = \tau$. It is then randomly restarted with new parameters.

Having decided on the shape of our plumes and their temporal dependence, we are now left to apply plumes to our boundary fields $\mathcal{R}$ and $\mathcal{S}$. Each plume is assigned an angular position $\vec{\gamma}_i$, an amplitude in both radial velocity $\mathcal{V}_i$ and entropy $\mathcal{E}_i$, an angular width $\delta_i$, and a lifetime $\tau_i$. With all of these parameters, the boundary fields are given by
\begin{equation}
\mathcal{R} \left( \theta, \phi, t \right) = \sum_{i=1}^{N_p} \mathcal{V}_i \, \mathcal{P} \left( \frac{ \Delta \sigma  \left(\vec{\gamma}, \vec{\gamma}_i \right)}{ \delta_i } \right) \, \mathcal{T} \left( t, t_{0i}, \tau_i \right)
\end{equation}
\begin{equation}
\mathcal{S} \left( \theta, \phi, t \right) = \sum_{i=1}^{N_p} \mathcal{E}_i \, \mathcal{P} \left( \frac{ \Delta \sigma  \left(\vec{\gamma}, \vec{\gamma}_i \right)}{ \delta_i } \right) \, \mathcal{T} \left( t, t_{0i}, \tau_i \right) .
\end{equation}
Here we have chosen to use the same widths for both the momentum and entropy profiles.

In the case of real near-surface plumes we can reasonably expect that plumes should be advected by large-scale flows. Indeed one of the claimed detections of giant cells relies on the advection of super-granules \cite{Hathaway2012}. In this work we have chosen a highly simplified treatment of plume advection where plume centers travel only in longitude at the rate of the axisymmetric differential rotation established by the simulation itself. We ignore any advection in latitude as well as local  (non-axisymmetric) longitudinal motions. This likely serves to suppress the meridional circulation and does not permit the aggregation of plumes, which would be over-estimated as a consequence of our plumes' very long lifetimes compared to those of supergranules. We anticipate adding full advection when plume lifetimes can be significantly shortened in future simulations with higher spatial and temporal resolution.

We now turn to the other two boundary conditions used in this formulation. Due to the nature of the anelastic equations solved by ASH, the introduction of plumes can generate significant pressure perturbations, which are then instantly transmitted to the entire domain. We minimize these pressure perturbations by choosing to make the plumes horizontally converging as they enter the domain as required by the spherical geometry and the stratification. Specifically, we choose 
\begin{equation}
\nabla_\perp \cdot \vec{u} = \left( \frac{ d \ln \bar{\rho} }{ d r } + \frac{2}{3} \frac{ d \ln \nu }{ d r } + \frac{ 2 }{ r } \right) u_r .
\label{eq:diverg}
\end{equation}
This is chosen so that when the horizontal divergence of the momentum equation is taken the largest terms on the right hand side cancel out.

Finally, with the horizontal divergence of the velocity field specified, we can still specify a condition on the toroidal component (any streamline that closes on a horizontal surface) of the horizontal velocities themselves. We do this using the standard stress-free boundary condition given by
\begin{equation}
\frac{ \partial }{ \partial r } \left( \frac{ u_\theta }{ r } \right) = \frac{ \partial }{ \partial r } \left( \frac{ u_\phi }{ r } \right) = 0
\label{eq:stress}
\end{equation}
applied only to the toroidal component of the horizontal velocities.

ASH employs a streamfunction decomposition for the mass flux in order to assure that the anelastic continuity equation is satisfied, with
\begin{equation} 
\bar{\rho} \vec{v} = \nabla \times \left( Z \hat{r} \right) + \nabla \times \nabla \times \left( W \hat{r} \right).
\end{equation} 
$W$ is thus the poloidal mass flux streamfunction and $Z$ the toroidal mass flux streamfucntion. In this formulation the boundary conditions given by \ref{eq:vr} and \ref{eq:diverg} are applied to the poloidal mass flux streamfunction $W$, and the boundary condition given by \ref{eq:stress} is applied to the toroidal mass flux streamfunction $Z$.

\end{appendix}

\end{document}